\documentclass[twocolumn,twocolappendix]{aastex631}
\usepackage{graphicx,graphics,amssymb}

\newcommand{\ter}{Terzan\,6}
\newcommand{\grs}{GRS\,1747--312}
\newcommand{\chandra}{{\em Chandra}}
\newcommand{\suzaku}{{\em Suzaku}}
\newcommand{\lum}{erg\,s$^{-1}$}

\newcommand{\dotsec}{$\rlap{.}^{\rm s}$}

\received{2023 December 29}
\revised{2024 February 19}
\accepted{2024 February 26}
\published{2024 May 9}
\shorttitle{Terzan\,6}
\shortauthors{van den Berg et al.}

\begin{document}

\title{Discovery of a second eclipsing, bursting neutron-star low-mass X-ray binary in the globular cluster Terzan\,6}

\correspondingauthor{Maureen van den Berg}
\email{mvandenberg@cfa.harvard.edu}

\author[0000-0003-0746-795X]{Maureen van den Berg}
\affiliation{Center for Astrophysics $\mid$ Harvard \& Smithsonian,
60 Garden Street,
Cambridge, MA 02138, USA}

\author[0000-0001-8371-2713]{Jeroen Homan}
\affiliation{Eureka Scientific, Inc., 2452 Delmer Street, Oakland, CA 94602, USA}

\author[0000-0003-3944-6109]{Craig O. Heinke}
\affiliation{Department of Physics, University of Alberta, CCIS 4-183, Edmonton, AB, T6G 2E1, Canada}

\author[0000-0003-4897-7833]{David A. Pooley}
\affiliation{Department of Physics and Astronomy, Trinity University, San Antonio, TX, USA}

\author[0000-0002-3516-2152]{Rudy Wijnands}
\affiliation{Anton Pannekoek Institute for Astronomy, University of Amsterdam, Postbus 94249, 1090 GE Amsterdam, The Netherlands}

\author[0000-0003-2506-6041]{Arash Bahramian}
\affiliation{International Centre for Radio Astronomy Research, Curtin University, Bentley, WA 6102, Australia}

\author[0000-0003-3124-2814]{James C.A. Miller-Jones}
\affiliation{International Centre for Radio Astronomy Research, Curtin University, Bentley, WA 6102, Australia}











\begin{abstract}
 We have analyzed {\it Chandra} and \suzaku\ observations of the globular cluster \ter, 
 made when the recurrent transient \grs\ was in quiescence. Our analysis reveals the presence of a second eclipsing, bursting neutron-star low-mass X-ray binary in the central regions of the cluster, in addition to \grs. The new source, which we name \ter\ X2, is located only $\sim$0.7\arcsec~away from \grs\  in the 2021 \chandra\ images. The detection of a 5.14 ks-long eclipse in the light curve of X2 at a time not predicted by the ephemeris of \grs\ confirms that it is an unrelated source. Using the \suzaku\ light curve from 2009, which in addition to a type-I X-ray burst also showed an eclipse-like feature, we constrain the orbital period to be longer than 16.27 hr. The 0.5--10 keV luminosities of X2 vary in the range of $\sim$0.24--5.9$\times10^{34}$ \lum\ on time scales of months to years.
We have identified a plausible optical counterpart of X2 in {\em HST} F606W and F814W images. This star varied by 2.7 mag in $V_{\rm 606}$ between epochs separated by years. In the cluster color-magnitude diagram, the variable counterpart lies in the blue-straggler region when it was optically bright, about 1.1--1.7 mag above the main-sequence turn-off. From the orbital period–density relation of Roche-lobe filling stars we find the mass-donor radius to be  $\gtrsim0.8$ $R_{\odot}$.

\end{abstract}

\keywords{}


\section{Introduction}

Terzan\,6 is a core-collapsed globular cluster at a distance of 6.7
kpc \citep{valeea07}. It is located in the direction of the bulge at only
$\sim$2$^{\circ}$ from the Galactic plane. As a result, Terzan\,6
suffers from high foreground extinction ($E(B-V)=2.35$; \citealt{valeea07}). The stellar encounter rate in Terzan\,6 is among
the highest of all Galactic globular clusters \citep{bahrea13}, and
consequently the number of products of such encounters, low-mass X-ray binaries (LMXBs) and
other compact binaries, is expected to be enhanced \citep{verbhut87}. In optical and 
near-infrared images, the appearance of Terzan\,6 is very compact: the 
core radius $r_c$ is 3\arcsec~and the half-light radius $r_h$ is 0.4\arcmin~
\citep{harr962010}.

Terzan\,6 is home to the long-known recurrent X-ray transient GRS\,1747$-$312,
which was discovered as a new, bright ($L_X\approx3\times10^{36}$ erg
s$^{-1}$, 0.5--2 keV) source in {\em ROSAT} and {\em GRANAT}
observations taken in 1990 September \citep{predea91,pavlea94}. 
As in almost all 
known persistent and transient Galactic globular-cluster LMXBs, the
compact object in GRS\,1747$-$312 is a neutron star on account of the
detection of thermonuclear flashes (type-I X-ray bursts) from the
neutron-star surface \citep{kuulea03,intzea03}. GRS\,1747$-$312 goes
into outburst quite regularly with an average recurrence time of 136
days \citep{intzea03,simo09}, with outbursts lasting about a
month. During outbursts, the light curves show total X-ray eclipses
with a period of 12.4 hr and a duration of 43 min \citep{intzea00,intzea03}. \cite{revnea02} measured the
position of GRS\,1747$-$312 using {\em Chandra X-ray Observatory}
High Resolution Camera imaging (HRC-I) data taken during outburst, the
accuracy of which was improved by \cite{intzea03}. The active source
at this epoch was confirmed to be GRS\,1747$-$312 with the detection
of eclipses in {\em RXTE}/PCA data taken during the same
outburst.

There are indications that \ter\ hosts another LMXB.  In {\it Suzaku} observations from September 2009 \citet{sajiea16} detected a source with a luminosity of $\sim$6.0$\times10^{34}$ erg\,s$^{-1}$ (for a 6.7 kpc distance), indicating that \grs\ was not in outburst, but perhaps in a low-luminosity state. {\it RXTE}/PCA bulge scan observations show that the most recent outburst of \grs\ had ended $\sim$60 days before. On two occasions during the {\it Suzaku} observation, the source did not show eclipse ingresses at the predicted times for \grs. The source also showed a type-I X-ray burst. The two missing eclipse ingresses suggest that {\it Suzaku} had detected another low-luminosity neutron-star LMXB in \ter.

Obtaining a census of the X-ray source population of Terzan\,6 is challenging. {\em Chandra} imaging has a spatial resolution of 0.4\arcsec--0.5\arcsec, while the core radius is only 3\arcsec. Moreover, the X-ray emission of the core is often dominated by \grs. When this source is in outburst, it completely outshines any fainter sources that may be present: the detected photons from \grs\ spread over the entire core radius, and the severely piled-up core of the point spread function (PSF) in ACIS images is very hard to correct for.

To look deeper into the core of Terzan\,6 in X-rays, we obtained observations of the cluster at times when \grs\ was in quiescence. We complement our observations with new and archival {\em Chandra} data taken when \grs\ was in outburst, and with {\em Suzaku} and {\em Hubble Space Telescope} {\em (HST)} observations. Here we present the results of our analysis, which confirm the presence of a second neutron-star LMXB in \ter. We show that, like \grs, this new source is eclipsing in X-rays, and we provide a lower-limit on the orbital period of the system. 

Sections \ref{sec_obs} and \ref{sec_ana} describe the observations and
analysis. In Section \ref{sec_res}, we present our results on the
identification and characterization of the new X-ray transient and our search for an optical
counterpart.  Finally, in Section \ref{sec_dis} we consider its nature and discuss reports in the literature where possible previous detections of the new source were (tentatively) attributed to \grs.

\vspace{1cm}
\section{Observations} \label{sec_obs}

\begin{small}
\begin{table*}
  \caption{{\em Chandra}, {\em Suzaku}, and {\em HST} observations of Terzan\,6 analyzed in this work \label{tab_obs}}
  \begin{center}
    \begin{tabular}{lll@{\hskip0.7cm}ll@{\hskip0.7cm}l@{\hskip0.7cm}l@{\hskip0.3cm}cc}
      \tableline \tableline
      Epoch & X-ray state     & Date      & \multicolumn{2}{l}{Instrument} & ObsID  & Exposure & \multicolumn{2}{c}{X-ray eclipse$^b$} \\
            &  GRS            &           &            &                   &        & time$^a$     & \multicolumn{1}{c}{Observed?}   & \multicolumn{1}{c}{Consistent?}   \\
            &           &           &            &                   &        &          &            &  \\
      \tableline
      1 & outburst   & 2000 Mar 9  & {\em Chandra} & HRC-I           & 720  & 9.97 ks  & no         & yes   \\
      \tableline

       2 & outburst & 2004 Mar 29 & {\em Chandra} & HETG$^c$             & 4551  &  45.03 ks  &  yes   & yes \\
      \tableline

       3 & quiescence$^d$ & 2009 Sep 16 & {\em Suzaku} & XIS             & 504092010  & 50.91 ks  &  no   & no \\
       \ldots & \ldots & \ldots  & \ldots & \ldots & \ldots & \ldots & yes & no \\     
       \ldots & \ldots & \ldots  & \ldots& \ldots& \ldots& \ldots & no & no \\     
      \tableline

      4 & quiescence & 2016 Jul 24 & {\em HST} & WFC             & 14074  & 2036 s (F606W) &  \ldots   &  \ldots \\
      \tableline
      5 & outburst & 2019 Jun 12 & {\em Chandra} & ACIS-S            & 21218  & 10.20 ks  &  yes       & yes \\
      \ldots   &  \ldots         & 2019 Jun 24 & {\em HST} & WFC               & 15616  & 1474 s (F606W)  &  \dots     & \ldots \\
      \ldots   &   \ldots        & \ldots      &  \ldots   &  \ldots           & \ldots &  1530 s (F814W) & \ldots & \ldots \\
      \tableline
      6 & quiescence & 2021 Apr 20 & {\em Chandra} & ACIS-S            & 23443  & 29.58 ks  & no        & no \\
       \ldots &  \ldots          &   \ldots          &    \ldots           &   \ldots                &   \ldots     &   \ldots        & yes & no \\
      \tableline
      7 & outburst & 2021 Jun 1 & {\em Chandra} & ACIS-S         & 23444  & 29.68 ks  & yes        & yes  \\
      \tableline
      8 & quiescence & 2021 Aug 23 & {\em Chandra} & ACIS-S            & 23441  &  9.82 ks  & no         & no  \\
      \ldots   & \ldots            & 2021 Sep 11 & {\em HST} & WFC             & 16420  &  1950 s (F606W) & \ldots     & \ldots \\
      \ldots   & \ldots            &   \ldots           &  \ldots           &    \ldots                  &  \ldots      &  1922 s (F814W) & \ldots & \ldots \\
      \tableline
    \end{tabular}
    \end{center}
    $^a$ For {\em HST} observations, we add the filter in which the exposure time was obtained between parentheses. $^b$ Was an X-ray eclipse observed, and was the  presence/absence of an eclipse consistent with the ephemeris of \grs\ from \cite{intzea03}? The consecutive rows for epochs 3 and 6 refer to multiple expected and unexpected eclipses; see Figures \ref{fig:chandra_lc} and \ref{fig:suzaku}.  $^c$ Data were obtained in continuous clocking mode, which does not provide 2D imaging. $^d$ See Section~\ref{sec:res_eclipses}.
\end{table*}
\end{small}

We have analyzed six {\em Chandra}, one {\em Suzaku}, and three {\em HST} observations of Terzan\,6, which are summarized in chronological order in Table~\ref{tab_obs}. In the table, we have also indicated the X-ray state of GRS\,1747$-$312 (i.e., outburst or quiescence) at the time of the observation, whether an X-ray eclipse was observed, and whether the presence/absence of eclipses was expected based on the ephemeris of \grs\ from \citet{intzea03}.

The data were acquired in 8 epochs. We obtained two contemporaneous {\em Chandra} and {\em HST} data sets
of Terzan\,6: once when GRS\,1747$-$312 was in outburst (epoch 5) and once when \grs\ was in quiescence (epoch 8). A few months prior to epoch 8, we observed the cluster with two deep {\em Chandra} observations separated by about 1.5 months, when GRS\,1747$-$312 was quiescent (epoch 6) and in outburst (epoch 7). For this work, we have complemented our observations with archival data from {\em Chandra} (epochs 1 and 2), {\em Suzaku} (epoch 3), and {\em HST} (epoch 4). The following sections provide more details about these observations, while the section headings indicate the X-ray state of \grs.

\subsection{Epoch 1: March 2000 outburst -- Chandra}\label{sec_chandra1}

On 2000 March 9 starting at 05:44 UTC, a 9.97 ks {\em Chandra} HRC-I observation of Terzan\,6 was performed when GRS\,1747$-$312 was in outburst (ObsID 720, P.I.\,Grindlay). 
There is no doubt about the identity of the bright X-ray transient in this observation. No eclipses are present in the HRC light curve, but as reported by \cite{intzea03},  {\em RXTE} observations of the same outburst clearly show eclipses that are consistent with the ephemeris
of GRS\,1747$-$312.

The HRC-I has a spatial
resolution of $\sim$0.4\arcsec\,sampled with 0.1318\arcsec\,pixels. This is slightly better than the spatial resolution of
the {\em Chandra} ACIS instrument ($\sim$0.49\arcsec) with which the
other Terzan\,6 observations that we use, were taken. Moreover, since the HRC-I
detector is a microchannel plate, the images of very bright sources do not suffer from the pile-up effects that occur in the ACIS CCD detectors. Therefore, \cite{revnea02} and
\cite{intzea03} were able to measure an accurate position for
GRS\,1747$-$312 from these HRC-I data.

\subsection{Epoch 2: March 2004 outburst -- Chandra}\label{sec_chandra2}

On 2004 March 29 starting at 22:53 UTC, {\em Chandra} observed \ter\ with HETG/ACIS-S for 45.03 ks (ObsID 4551, P.I. M\'endez). ACIS-S was operated in continuous-clocking mode, which only provides spatial information in one dimension. Data from the {\em RXTE}/PCA Bulge Scan Program\footnote{\url{https://asd.gsfc.nasa.gov/Craig.Markwardt/galscan/html/GRS_1747-312.html}} indicate that a source was in outburst in \ter, and its occurrence was consistent with the pattern of outbursts from \grs~\citep{simo09}.

\subsection{Epoch 3: September 2009 quiescence -- Suzaku} \label{sec_suz}

A {\em Suzaku} XIS observation was performed on 2009 September 16 (ObsID 504092010, P.I Koyama). It started at 07:22 UTC, had a duration of $\sim120$ ks and an effective exposure of 50.91 ks. {\em RXTE}/PCA Bulge Scan data indicate that no source was in outburst at the time of the {\em Suzaku} observation\footnote{A large flare can be seen in the {\em RXTE}/PCA Bulge Scan data four days prior, but the detection of pulsations at $\sim$245 Hz suggest that this flare was due to contamination from an ongoing outburst of IGR J17511–305 \citep{markwardt2009}.}. The spatial resolution of {\em Suzaku}'s X-ray telescopes (2\arcmin) is not sufficient to resolve the X-ray source population in the core of \ter, but, as we  show in Section \ref{sec:res_eclipses}, \grs\ was truly in quiescence. A detailed analysis of the {\it Suzaku} data can be found in \citet{sajiea16}.

\subsection{Epoch 4: June 2016 quiescence -- HST} \label{sec_hst1}

Prior to our observations of the 2019 outburst of GRS\,1747$-$312 (Section \ref{sec_hst2}), Terzan\,6 was observed only once with {\em
  HST}. Under program GO\,14074 (P.I.\,Cohen), five images in the
F606W filter (1$\times$60 s, 4$\times$494 s) were taken with the
Advanced Camera for Surveys (ACS) Wide Field Channel (WFC) on 2016
July 24 starting 05:45 UT. Long-term 
  MAXI light curves\footnote{\url{http://maxi.riken.jp/star_data/J1750-312/J1750-312.html}} indicate that no source was in outburst (above $10^{36}$ \lum) in \ter\ at the
time of this {\em HST} observation. The 3.4\arcmin~$\times$
3.4\arcmin~WFC field of view is covered by two CCDs with a 0.05\arcsec~pixel scale, separated by a
50-pixel (2.5\arcsec) chip gap. The core of the cluster was placed in
the center of this field. Since the long exposures were dithered with
offsets that span the chip gap (up to 12\arcsec), the combined images
fully cover the central parts of the cluster.

The GO\,14074 data set also includes F110W and F160W images taken with
the Wide Field Camera 3 Infra Red Channel (WFC3-IR). The WFC3-IR
detector resolution (0.13\arcsec~pixel$^{-1}$) is lower than that of
ACS/WFC. Since the coarser sampling complicates source detection and the extraction of
photometry in the crowded cluster core even more, we did not analyze
these exposures.

\subsection{Epoch 5: June 2019 outburst -- Chandra and HST} \label{sec_hst2}

{\em Swift} monitoring of the Galactic Bulge (P.I. Maccarone) and \ter\ (P.I. Heinke) indicated that in early June 2019 a source in the cluster had become active. Subsequently, we triggered our {\em Chandra} and {\em HST} Target of
Opportunity (ToO) program to localize and identify transient LMXBs in Galactic globular clusters. We
observed Terzan\,6 for a total of 10.20 ks with the ACIS-S array
starting 2019 June 12 13:31 UTC (ObsID 21218). {\it Swift} and {\it MAXI} observations indicate that the observation took place near the peak of the outburst. Data were taken in
faint, timed-exposure mode with GRS\,1747$-$312 placed near the
aimpoint on the S3 chip but shifted by 1.8\arcmin~in the Z direction (i.e.~along the short side of the array). This Z-offset, together with the activation of two more CCDs (S2 and and S4), allowed for the detection of more
sources outside the cluster that could be used for registering this image to other {\em Chandra}
observations and to optical source catalogs. 

Twelve days later, on 2019 June 24 at 06:01 UT, we observed Terzan\,6 with the ACS/WFC (program
GO\,15616). To optimize the
match with the ACS/WFC data from GO\,10474 (Section \ref{sec_hst1}),
we took images in the F606W filter. As we wanted to measure or constrain colors, we
also took data in a second filter. Since Terzan\,6 is highly reddened,
we opted for F814W. The data were taken in a single visit, with one
orbit dedicated to imaging in the F606W filter, and a second orbit to
F814W imaging. In each filter, one short (30 s) and four long
exposures were taken (4$\times$361 s in F606W, 4$\times$375 s in
F814W). The long exposures were arranged in a 4-point dither pattern
with small fractional-pixel offsets which allowed us to slightly
improve the image resolution in the stacked images compared to the
detector scale of 0.05\arcsec~per pixel. These offsets are too small
to span the chip gap, and therefore the core of the cluster was
centered on one of the chips to fully cover it with the WFC images.

\subsection{Epoch 6: April 2021 quiescence -- Chandra}\label{sec_chandra4}

We obtained a 29.58 ks ACIS-S image of Terzan\,6 on 2021 April 20
starting 20:11 UTC (ObsID 23443). The goal of this observation was to
probe the population of faint X-ray sources in the cluster. This required
that no bright cluster X-ray sources were present at the time of the
observation. An accompanying {\em Swift} program, comprising a sequence
of 1 ks observations taken with a cadence of several days, was in place to
ensure this was indeed the case. The ACIS-S data were taken in very
faint, timed-exposure mode with six active CCDs (S1 to S4, I2
and I3) and with GRS\,1747$-$312 placed near the S3 aimpoint.

\subsection{Epoch 7: June 2021 outburst -- Chandra}\label{sec_chandra5}

We obtained a second ACIS-S observation complementary to ObsID 23443
to study faint X-ray sources in Terzan\,6. However, inspection of
the 29.68 ks exposure obtained on 2021 June 1 starting 11:28 UTC (ObsID
23444) revealed that GRS\,1747$-$312 had become active
again, reducing the sensitivity to detecting faint sources in a large part
of the area within the cluster half-light radius. The observation parameters
were the same as for ObsID 23443 except that the only active CCDs were
S2, S3 and S4.  {\it Swift} monitoring observations (PI: Motta) that started a week after our \chandra\ observation, suggest that the ACIS-S observation took place during the outburst rise. 

\subsection{Epoch 8: August/September 2021 quiescence -- Chandra and HST} \label{sec_chandra6}

In mid-August 2021 we again triggered our ToO program after {\em
  Swift} observations (P.I. Bahramian) detected activity of a possible transient in Terzan\,6. The 9.82 ks ACIS-S
observation (ObsID 23441) was taken on 2021 August 23 starting 09:53
UTC in faint, timed-exposure mode with GRS\,1747$-$312 near the
aimpoint. By the time of this observation the transient had already faded, and no signs of activity from GRS\,1747$-$312 or any other transients are visible in the ACIS-S
images anymore. The active CCDs during this observation are S1 to
S4, I2 and I3.

On 2021 September 11 starting 07:11 UT, we obtained F606W (30 s,
4$\times$480 s) and F814W (30 s, 4$\times$473 s) ACS/WFC images of
Terzan\,6. The observation setup was very similar to the one
adopted in Epoch 4 for GO15616 (Section\ref{sec_hst2}).

\begin{figure*}
  \centerline{\includegraphics[width=18cm]{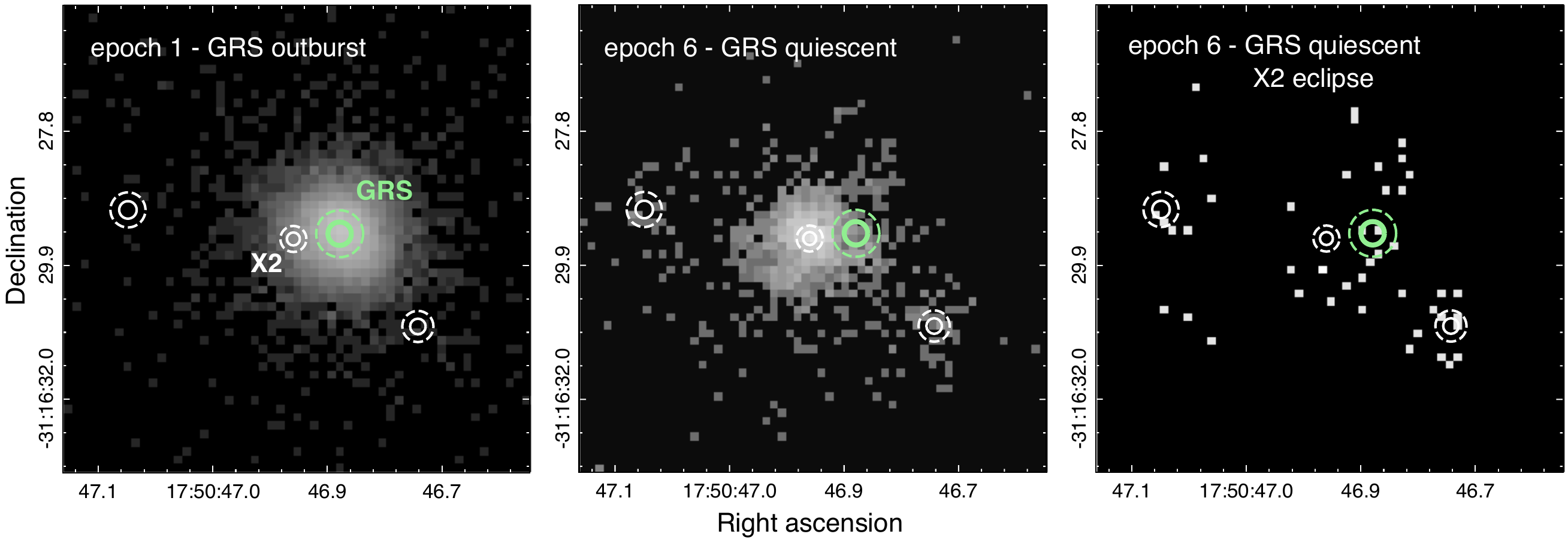}}
    \caption{{\em Chandra} images of the core of \ter\ from epoch 1 (left; HRC-I, 0.8--10 keV) when \grs\ was in outburst, and epoch 6 (middle, right; ACIS-S, 1--7 keV) when \grs\ was in quiescence. All images are aligned to the {\em HST} astrometry. The detection in the HRC-I image is plotted with a green circle in all panels; sources detected in the image shown in the middle panel are plotted with white circles in all panels. The 1 and 2 $\sigma$ errors are shown with solid and dashed circles.
     The middle panel is an image from the entire observation ObsID 23443, where X2 is the brightest source. The right panel only shows ObsID-23443 events from when X2 was in eclipse, revealing a number of fainter nearby sources. The ACIS-S images use 1/4 pixel sub-resolution. North is up, east is to the left. \label{fig_chandra}}
\end{figure*}

\section{Data reduction and analysis} \label{sec_ana}

\subsection{Chandra}

\subsubsection{Image analysis}

All {\em Chandra} data were reprocessed and analyzed with CIAO 4.15.2 in
combination with CALDB 4.10.7. We created light curves from large
source-free regions to check for background flares, but did not find
any variations larger than 3 $\sigma$ from the average background
count rate. In all imaging observations, multiple sources are clearly visible
outside the cluster, which were used to register the {\em Chandra}
images to each other and to the {\em HST} astrometry.

ObsIDs 21218 and 23444 were performed when GRS\,1747$-$312 was in outburst 
(see Section \ref{sec_hst2} and \ref{sec_chandra5}), which resulted in a heavily piled-up 
doughnut-shaped image but also in strong readout streaks. To avoid spurious source 
detections, the readout streak was removed with the task {\tt acisreadcorr} before 
performing the source detections. Source positions were measured by running {\tt wavdetect} on 
all ACIS images in a wide band (1--7 keV) and a soft band (0.5--1.2 keV), with a significance threshold of $10^{-6}$ and wavelet scales of 
1, 2, 4, 8, and 16 pixels. The resulting sources were visually inspected in the images. 
Spurious detections in the piled-up core 
were removed from the lists. For each observation, the resulting source lists for the 
two energy bands were matched to create a master source list. The visual inspection also 
revealed that in ObsID 23443, which was taken when there were no active transients in Terzan\,6, 
{\tt wavdetect} missed a few faint sources near the brightest source in the cluster core. 
To remedy this, we created images of these observations with a binfactor of 0.25 and reran 
{\tt wavdetect} with scales 1, 2, and 4, after which these faint sources were successfully 
picked up.

HRC data have poor energy resolution so for ObsID 720 we ran {\tt wavdetect} without 
applying an energy filter. 

We defer a discussion of the properties of all sources detected in Terzan\,6 to a follow-up paper 
and focus on the central, brightest, two sources here.

\subsubsection{Alignment of Chandra source catalogs} \label{sec_xxbore}

We aligned the source lists of the individual {\em Chandra} imaging observations to the same astrometric frame. We chose to register 
all observations to ObsID 23443. This observation is the deepest {\em Chandra} exposure that 
we analyzed in which the sensitivity in the core is not drastically reduced by a bright transient.
To determine the relative offset between an observation and ObsID 23443, we identified the detections
that both observations have in common and compute the mean offset in right ascension and declination.
We excluded sources with fewer than 10 net {\tt wavdetect} counts (1--7 keV) and with {\tt wavdetect} positional 
errors larger than 0.8\arcsec~in either direction. Inspection of ObsID 23443 shows that 
the cluster core contains several closely separated sources (see middle  panel of Fig.\ \ref{fig_chandra}). Some of 
these are not detected in the other observations, which makes it harder to securely cross-identify them between 
ObsIDs. Including mismatches in measuring the alignment would skew the resulting relative offset, and therefore we did not use sources 
inside the cluster half-light radius. 

Table~\ref{app_tab_xoffsets} in Appendix~\ref{app_xalign} summarizes the mean relative offsets between ObsID 23443 
and the other ObsIDs. In Table~\ref{app_tab_xcoords} we list the coordinates of the sources that we used to 
determine the offsets.

\subsubsection{Light curves}

Prior to our light curve extraction we barycentered all \chandra\ observations using the {\tt axbary} tool in CIAO. For the coordinates we used those of the brightest source in the image; for ObsID 4551 we used our coordinates of \grs\ from the HRC-I observation, since no accurate position could be determined. For each imaging observation we created light curves for the brightest source in the cluster; details for the non-imaging observation 4551 are provided below. Light curves were created with the CIAO task {\tt dmextract} using a variety of extraction region sizes and time resolutions. 

For the HRC-I observation (ObsID 720) we used a circular extraction region with a radius of 1.5\arcsec~and a time resolution of 50\,s. Since the HRC-I has a poor energy resolution, the light curve was created for the full HRC-I energy band (0.8--10 keV). ACIS-S light curves were created in the 0.3--10 keV energy band. For ObsIDs 21218 and 23444, made when \grs\ was in outburst, we used extraction regions with radii of 7\arcsec~and a time resolution of 50\,s. For ObsIDs 23441 and 23443, made when \grs\ was in quiescence, we used extraction radii of just 0.5\arcsec, to avoid contamination from nearby faint sources (see Fig.~\ref{fig_chandra}); given the lower count rates, a time resolution of 150\,s was used. 

For the light curve of ObsID 4551, which was made with the HETG in continuous-clocking mode, we used an extraction region of 6\arcsec~wide, centered on the brightest pixel in the 1D image. The energy band was 0.3--10 keV and the time resolution 50\,s.

\subsubsection{Spectra}

A spectral analysis was performed for the two observations during which \grs\ was in quiescence (ObsIDs 23441 and 23443) and for the segment during which \grs\ was in eclipse in outburst observation 23444. In these observations/segments the flux is likely dominated by a source that is not \grs\ (see Sections \ref{sec:res_imaging} and \ref{sec:res_eclipses}). We used the CIAO tool {\tt specextract} to extract the spectra. The new source is surrounded by several faint sources. For observations 23444 and 23441, we therefore used small (0.5\arcsec\,radius) circular extraction regions for the source spectra, to minimize contamination, and source-free circles with radii of 5\arcsec\, for the background spectra. During observation 23443, the new source showed a full eclipse (see Fig.~\ref{fig:chandra_lc}d), allowing us to use the eclipse segment for the background spectrum. For both the source spectrum (excluding the eclipse) and the background spectrum we used the same 1\arcsec-radius extraction circle. The resulting spectra were grouped to a signal-to-noise of 4.5 with {\tt dmgroup} and fit with XSPEC 12.13.0 \citep{arnaud1996} in the 0.5--10 keV range using $\chi^2$-statistics. Abundances were set to {\em wilm} \citep{wilms2000} and cross-sections to {\em vern} \citep{vern1996}.

\subsection{HST}

The ACS calibration pipeline produces calibrated flat-fielded images
that are corrected for charge transfer efficiency losses. We retrieved
these {\tt flc} images of programs GO\,14074, GO\,15616 and GO\,16420
from the STScI archive and used them as the starting point for our
analysis. For each program, we combined the individual images of each
filter into a deeper and oversampled stacked image that is corrected
for geometric distortion using the {\tt astrodrizzle} task in the DrizzlePac
package. The short (60\,s for GO\,14074, 30\,s for GO\,15616 and
GO\,16420) exposures, which were primarily aimed at obtaining
photometry for the bright stars in the field, were excluded from the
stacks. In our stacked images the pixel scale is smaller by a factor
$\sim$1.5 (0.033\arcsec~per pixel) than in the individual images, and
effects of cosmic rays, hot pixels and bad columns are significantly
reduced. We derived an astrometric solution for the stacked images by
fitting the pixel positions of around 2000 detections of Gaia DR2
stars to their catalogued positions (epoch 2015.5;
\citealt{gaiadr218}). Fitting for pixel scale, rotation and zeropoint
gives rms residuals of $\lesssim$0.012\arcsec~in each direction.

We used the  Dolphot2.0\footnote{\url{http://americano.dolphinsim.com/dolphot/}; see
  also \cite{dolp00}} suite of routines to extract photometry from the
{\tt flc} images. The {\tt Dolphot} photometry is calibrated to the
Vegamag scale using the WFC zeropoints and encircled energy
corrections from \cite{bohl12}. In this paper, we write calibrated magnitudes in the F606W and F814W filters as $V_{\rm 606}$ and $I_{\rm 814}$, respectively. We followed the recommended parameter
settings from \cite{coheea18}, who performed extensive tests on the
optimal performance of {\tt Dolphot} for all ACS/WFC images in
their GO\,14074 program. This program targeted several globular clusters; their data for Terzan\,6 are the images for our epoch 4. 
\cite{coheea18} find that faintward of the 50\% completeness limit the photometry is not
trustworthy. For Terzan\,6, this limit corresponds to $V_{\rm
  606}\approx 26.9$ (see Table 4 in \cite{coheea18}).

We compared the F606W and F814W magnitudes of stars between the epochs to check for systematic differences. The magnitude offset between the GO\,15616 and GO\,16420 photometry is very small ($<$0.03$\pm$0.06) in both filters, whereas there is a small systematic magnitude offset in the F606W photometry of GO\,15616 and GO\,14074 ({\tt Dolphot} magnitudes in GO\,15616 are $\sim$0.16$\pm$0.1 brighter), which we corrected for.

\begin{deluxetable}{lllll}
\tabletypesize{\scriptsize}
\tablecaption{X-ray coordinates (RA, Dec) of the sources used to compute the X-ray/optical boresights. Positions are aligned to the {\em HST} frame (Gaia DR2; epoch 2015.5). Errors on the right ascension (e\_RA) and declination (e\_Dec) are the {\tt wavdetect} positional errors (in units of arcseconds). The last column is the identification number of the matching Gaia DR2 star.\label{tab_xopt_bore}}
\tablehead{RA & e\_RA & Dec & e\_Dec & Gaia DR2 ID \\ & (\arcsec) & & (\arcsec) & }
\startdata
\tableline
\multicolumn{5}{l}{ObsID 720}\\
\tableline
17$^{\rm h}$50$^{\rm d}$35\dotsec28  &  0.22    &        $-$31$^{\circ}$18\arcmin25\farcs9  &   0.2    &           4055719338235656448 \\
17$^{\rm h}$51$^{\rm d}$00\dotsec38  &  0.32    &        $-$31$^{\circ}$16\arcmin31\farcs5  &   0.3    &           4055718998983516672 \\
17$^{\rm h}$50$^{\rm d}$26\dotsec64  &  0.38    &        $-$31$^{\circ}$14\arcmin01\farcs4  &   0.3    &          4055731643367375488 \\
\tableline
\multicolumn{5}{l}{ObsID 23443}\\
\tableline
17$^{\rm h}$50$^{\rm d}$35\dotsec29 &   0.11     &       $-$31$^{\circ}$18\arcmin25\farcs6  &   0.1    &           4055719338235656448\\
17$^{\rm h}$50$^{\rm d}$32\dotsec65 &   0.13     &       $-$31$^{\circ}$17\arcmin10\farcs1  &   0.2    &           4055731196690716672\\
17$^{\rm h}$50$^{\rm d}$46\dotsec04 &   0.11     &       $-$31$^{\circ}$15\arcmin09\farcs8  &   0.1    &           4055719995415996672\\
17$^{\rm h}$51$^{\rm d}$00\dotsec40 &   0.20     &       $-$31$^{\circ}$16\arcmin31\farcs6  &   0.2    &           4055718998983516672\\
\tableline
\multicolumn{5}{l}{ObsID 23444}\\
\tableline
17$^{\rm h}$50$^{\rm d}$35\dotsec26 &   0.05     &       $-$31$^{\circ}$18\arcmin25\farcs8  &   0.1    &           4055719338235656448\\
17$^{\rm h}$50$^{\rm d}$32\dotsec65 &   0.10     &       $-$31$^{\circ}$17\arcmin10\farcs1  &   0.1    &           4055731196690716672\\
17$^{\rm h}$50$^{\rm d}$57\dotsec37 &   0.25     &       $-$31$^{\circ}$19\arcmin55\farcs3  &   0.2    &           4055718208681662336\\
17$^{\rm h}$51$^{\rm d}$00\dotsec38 &   0.20     &       $-$31$^{\circ}$16\arcmin31\farcs9  &   0.1    &           4055718998983516672\\
17$^{\rm h}$50$^{\rm d}$46\dotsec07 &   0.17     &       $-$31$^{\circ}$15\arcmin09\farcs5  &   0.2    &           4055719995415996672\\
\tableline
\multicolumn{5}{l}{ObsID 23441}\\
\tableline
17$^{\rm h}$50$^{\rm d}$35\dotsec28 &   0.11     &       $-$31$^{\circ}$18\arcmin25\farcs9  &   0.1    &           4055719338235656448\\
17$^{\rm h}$50$^{\rm d}$32\dotsec65 &   0.49     &       $-$31$^{\circ}$17\arcmin09\farcs8  &   0.1    &           4055731196690716672\\
\tableline
\multicolumn{5}{l}{ObsID 21218}\\
\tableline
17$^{\rm h}$50$^{\rm d}$35\dotsec28 &   0.23     &       $-$31$^{\circ}$18\arcmin25\farcs7  &   0.2    &           4055719338235656448\\
17$^{\rm h}$50$^{\rm d}$32\dotsec66 &   0.10     &       $-$31$^{\circ}$17\arcmin09\farcs9  &   0.2    &           4055731196690716672\\
17$^{\rm h}$50$^{\rm d}$17\dotsec56 &   0.24     &       $-$31$^{\circ}$14\arcmin36\farcs8  &   0.3    &           4055732330562155648\\
17$^{\rm h}$50$^{\rm d}$26\dotsec61 &   0.25     &       $-$31$^{\circ}$14\arcmin01\farcs7  &   0.3    &           4055731643367375488\\
\enddata
\end{deluxetable}

\subsection{Alignment of the Chandra and HST catalogs}

In order to look for optical counterparts to the central bright sources in Terzan\,6, the {\em Chandra} and {\em HST} catalogs 
need to be aligned. The {\em HST} images are already registered to the Gaia DR2 catalog, which in turn provides positions that are tied to the International Celestial Reference System (ICRS). The {\em  Chandra} astrometry, however, can be systematically offset from the ICRS, with different offsets (boresight corrections) for each individual ObsID. The 90\% uncertainty radius on the absolute astrometry is $\sim$1.1\arcsec~for ACIS-S and HRC-I\footnote{\url{https://cxc.harvard.edu/cal/ASPECT/celmon/}}. An added benefit of aligning all {\em Chandra} observations to this common astrometric reference frame is that it provides another check of the relative positions of the sources in the core of Terzan\,6: different, but overlapping, sets of sources are used for the {\em Chandra} versus {\em Chandra} alignment (Section \ref{sec_xxbore}) and the {\em Chandra} versus {\em HST} alignment.

Terzan\,6 lies at low Galactic latitude and is obscured by a neutral hydrogen column density $N_{\rm H}\approx$ 1.5--1.9$\times10^{22}$ cm$^{-2}$ \citep{vatsea18b} and reddening $E(B-V)\approx$ 2.35. The X-ray emission of stars that lie further away from us is increasingly absorbed, which hardens the X-ray spectra (or X-ray colors) for more distant stars. The probability to find a true counterpart in the Gaia DR2 catalog is therefore higher for foreground (i.e.\,soft) X-ray sources.\footnote{The magnitude limit of the Gaia DR2 catalog depends on the region of the sky, and can be $G=21$ but also as bright as $G=18$ for crowded regions \citep{gaiadr218}.} For each {\em Chandra} imaging observation, we therefore first selected the sources detected in the soft band. The HRC source list was compared with the ACIS source lists to check which HRC detections are soft sources. Subsequently, we matched these against stars in the Gaia DR2 catalog with the CIAO task {\tt wcs\_match} to derive the translation required to align the {\em Chandra} astrometry to the {\em HST} astrometry. We looked for initial matches inside a search radius of 1.5\arcsec~and in this step omitted sources with multiple Gaia stars inside this radius. In all cases, a clear initial shift emerged from the resulting X-ray/optical pairs. Sources that matched with two (which was the maximum we found) Gaia stars were then examined, and the X-ray/optical pair with an offset consistent with the initial shift was allowed back in, to compute the final shift. The task {\tt wcs\_update} was used to apply the offsets to the {\em Chandra} images and sourcelists. 

Since the field of view of ACS images is only 3.4\arcmin$\times$3.4\arcmin, typically zero or one soft source (not including sources in the core of \ter) falls inside. Therefore, we cannot derive the X-ray/optical boresights directly from matches between soft sources and stars in the {\em HST} images.

\begin{deluxetable}{rrrrr}
\tabletypesize{\scriptsize}
\tablecaption{Boresight offsets between the {\em Chandra} images and the Gaia DR2 catalog. Offsets are defined as $\Delta$RA=RA$_{\rm X}$ $-$ RA$_{\rm Gaia}$, $\Delta$Dec=Dec$_{\rm X}$ $-$ Dec$_{\rm Gaia}$. Offsets, and errors on the offsets, are given in units of arcseconds.\label{tab_xobore}}
\tablehead{\multicolumn{1}{c}{RA} & \multicolumn{1}{c}{e\_RA} & \multicolumn{1}{c}{Dec} & \multicolumn{1}{c}{e\_Dec} & \multicolumn{1}{c}{Gaia DR2 ID}}
\tablehead{ObsID & $\Delta$RA & e\_$\Delta$RA & $\Delta$Dec & e\_$\Delta$Dec \\ & (\arcsec) & (\arcsec) & (\arcsec) & (\arcsec) }
\startdata
\tableline
720    &  $-$0.010    & 0.15  &   $-$0.14    &  0.10 \\
23443  &     0.32     & 0.08  &      0.86    &  0.06 \\
23444  &     0.57     & 0.09  &      0.60    &  0.07 \\ 
23441  &  $-$0.52     & 0.13  &   $-$0.94    &  0.13 \\
21218  &  $-$0.22     & 0.04  &      0.29    &  0.08 \\
\tableline
\enddata
\end{deluxetable}

In Table~\ref{tab_xopt_bore} we provide the coordinates and {\tt wavdetect} positional uncertainties\footnote{For ACIS sources, alternative positional uncertainties can be computed with Eqs. 12--14 in \cite{kimea07}, which are formulae to compute the 95\%, 90\% or 68\% error radius ($r_{95}$ etc.) as a function of off-axis angle (from the observation aimpoint) and {\tt wavdetect} counts. We compared the {\tt wavdetect} errors (e\_RA and e\_Dec added in quadrature) with the Kim errors ($r_{95}$/2.45, to convert to a 1-$\sigma$ error) and found that, for all sources in our ACIS ObsIDs within 8\arcmin\ of the aimpoint and with fewer than 1000 counts, the ratio of ($r_{95}$/2.45) 
to 
the combined {\tt wavdetect} error is 0.9--1.1 on average.} of the X-ray/optical matches that we used for the alignment. Table~\ref{tab_xobore} lists the resulting boresight shifts $\Delta$RA and $\Delta$Dec. We compute the error on the boresight in RA and Dec (e\_$\Delta$RA, e\_$\Delta$Dec) as the root-mean-square of the residual (X-ray minus optical) positions after correcting for the boresight shift, divided by the square-root of the number of matches.

\subsection{Suzaku}

The single {\it Suzaku} observation of Terzan\,6 (see details in Table \ref{tab_obs}) was analyzed with the {\it Suzaku} FTOOLS that are part of HEASOFT v6.30.1. The data were reprocessed using {\tt aepipeline} with the latest calibration products. The event files of the XIS0, XIS1, and XIS3 detectors were barycenter corrected with {\tt aebarycen} and combined into a single event file using {\tt xselect}. We then created a 0.5--10 keV light curve with a time resolution of 80 s from a 75\arcsec-radius circular region around the source. No spectral analysis was performed; for spectral fits we refer to \citet{sajiea16}.

\section{Results} \label{sec_res}

\subsection{Terzan\,6 X2: a close neighbor of \grs}

Our analysis of the central X-ray emission from Terzan\,6 provides strong indications that the cluster hosts a second eclipsing neutron-star LMXB, which we name Terzan\,6 X2 (or X2), hereafter. There are two lines of results pointing to this: first, our astrometric analysis (Section \ref{sec:res_imaging}) and second, our light curve analysis (Section \ref{sec:res_eclipses}). The results of the X-ray spectral fitting of X2 and our search for optical counterparts to X2 and \grs\ are described in Sections \ref{sec:res_xspec} and \ref{sec:res_opt}, respectively.

\begin{deluxetable}{lllll}
\tabletypesize{\scriptsize}
\caption{Positions of \grs\ and X2 as measured from the different {\em Chandra} ObsIDs. The last column is the combined 1$\sigma$ error radius obtained by adding the {\tt wavdetect} and boresight errors in quadrature (given in units of arcseconds). \label{tab_xpos}}
\tablehead{Source & ObsID & RA & Dec & $\sigma$ (\arcsec)}
\startdata
\grs      & 720    & 17$^{\rm h}$50$^{\rm d}$46\dotsec848 & $-$31$^{\circ}$16\arcmin29\farcs44 & 0.18 \\
\ter\ X2  & 23443  & 17$^{\rm h}$50$^{\rm d}$46\dotsec905 & $-$31$^{\circ}$16\arcmin29\farcs52 & 0.10 \\
\ter\ X2  & 23441  & 17$^{\rm h}$50$^{\rm d}$46\dotsec914 & $-$31$^{\circ}$16\arcmin29\farcs66 & 0.19 \\ 
\ter\ X2  & 23444  & 17$^{\rm h}$50$^{\rm d}$46\dotsec901 & $-$31$^{\circ}$16\arcmin29\farcs74 & 0.15 \\
\grs?     & 21218  & 17$^{\rm h}$50$^{\rm d}$46\dotsec859 & $-$31$^{\circ}$16\arcmin29\farcs65 & 0.11 \\
\enddata
\end{deluxetable}

\begin{figure}
    \centering
    \includegraphics[width=8.5cm]{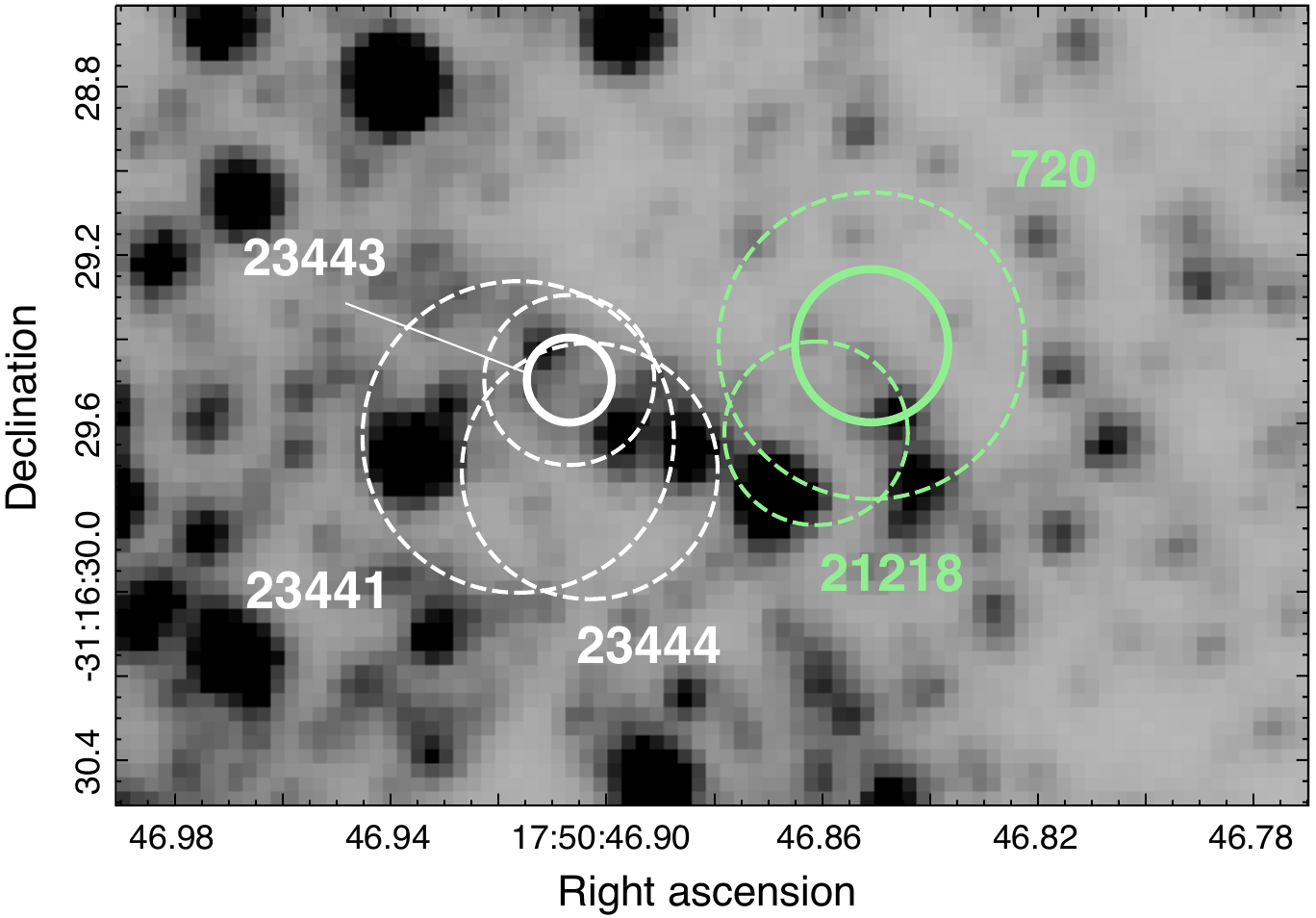}
    \caption{{\em HST} F606W image (2.85\arcsec~$\times$ 1.9\arcsec) from epoch 5. Overplotted in green are the {\em Chandra} position of the central source in Terzan\,6 from the HRC-I observation ObsID 720 (\grs\ in outburst) and from the \grs-eclipse interval in the ACIS-S observation ObsID 21218. The white circles mark the positions from the ACIS-S observations made when \grs\ was in quiescence (ObsIDs 23443 and 23441), and from the \grs-eclipse interval in ACIS-S observation ObsID 23444 (\grs\ in outburst, but $\sim$5$\times$ fainter than in 21218). Solid and dashed circles mark the 1 and 2 $\sigma$ error radii, respectively. The 1 $\sigma$ circles are only shown for ObsID 720 and 23443. North is up, east is to the left.}
    \label{fig:chandra_pos}
\end{figure}

\subsubsection{Astrometry} \label{sec:res_imaging}

There are two ways in which we obtained improved (i.e.\,aligned) astrometry for the X-ray sources in the core: registration of the {\em Chandra} images with respect to one another, and registration of each of these observations to the Gaia DR2 astrometry. These methods use different but overlapping sets of X-ray sources outside the cluster (at $r>r_h$), but give consistent results regarding the presence of a close and bright ($L_{\rm X} > 10^{33}$ erg s$^{-1}$) neighbor of \grs. Since the X-ray/optical alignment yields the smallest astrometric errors, we focus on the results of this method. In Appendix~\ref{app_xalign} we show the results of the alignment of the {\em Chandra} images against each other.

The results of aligning the {\em Chandra} positions with the Gaia DR2 ({\em HST}) astrometry can be seen in Fig.~\ref{fig_chandra}.  All images in the figure are aligned to the same reference frame. The left panel shows the HRC-I image from epoch 1 (\grs\ in outburst) while the middle panel shows an ACIS-S image from ObsID 23443 (epoch 6) when \grs\ was quiescent. The position of the source detected in the left panel is marked in green, while the positions of the sources detected in the middle panel are marked in white. The 1 and 2 $\sigma$ errors are shown with solid and dashed circles, where $\sigma$ is the quadratic sum of the {\tt wavdetect} error and the error on the X-ray/optical boresight offset.

Fig.~\ref{fig_chandra} shows that the position of \grs\, from ObsID 720 does not align with that of the brightest source, X2, in ObsID\,23443: the latter lies $\sim$0.7\arcsec~to the east of \grs. This is 3.4 times the error on the separation between the two sources, which we have calculated as the quadratic sum of the errors ($\sigma$) on each of the positions of \grs\ and X2. 
X2 dominates the X-ray emission from the core of \ter\ when \grs\ is in quiescence. Two other sources are detected at a separation of several arcseconds from X2 and the distribution of the events suggest some unresolved or diffuse emission, possibly originating from multiple distinct, faint sources. In Table~\ref{tab_xpos}, we report the {\em Chandra} positions of \grs~and X2.

Fig.~\ref{fig:chandra_pos} is an ACS F606W image from epoch 5 of the region around \grs\ and X2, where we show the same positions of those two sources as in Fig.~\ref{fig_chandra}. In addition, we show the position of the brightest central source in ObsID 23441, which is the other observation taken when \grs\, was in quiescence. This position is consistent with that of X2. 

The two other {\em Chandra} observations, ObsIDs 21218 and 23444, were taken when \grs\,was in outburst. We are confident that the bright source in these observations is \grs~because its light curve shows eclipses at the expected times for \grs~(see Section \ref{sec:res_eclipses}). A source position for \grs~can in principle be estimated by hand from the doughnut-shaped central PSF in these ObsIDs, 
but the (not piled-up) HRC-I detection of ObsID 720 provides an intrinsically better position. However, during both ObsID 21218 and 23444, \grs~went into eclipse, providing an opportunity to measure the position of \grs\ or of potentially other fainter X-ray sources in the core. This lasted $\sim$1910\,s for ObsID 21218 (as the eclipse occurred towards the end of the observation, see Fig.\,\ref{fig:chandra_lc}b) and $\sim$2460\,s in ObsID 23444 (Fig.\,\ref{fig:chandra_lc}c). Their positions could be measured by running {\tt wavdetect} on the eclipsed portions of these ObsIDs.

The boresight offsets measured from the entire ObsIDs 21218 and 23444 were applied to the resulting source lists and the result is included in Fig.~\ref{fig:chandra_pos}: the position of the brightest source seen when \grs\ is in eclipse aligns with X2 in ObsID 23444 but lies closer to \grs\ in ObsID 21218. We note that the eclipse count rate of ObsID 21218 was about five times higher than the eclipse count rate of ObsID 23444; the X-ray flux from \grs\ in eclipse (due to light scattered into our line of sight by an extended structure, such as the accretion disk corona) may have been high enough to outshine X2, whereas this may not have been the case in ObsID 23444.  Fig.~\ref{app_fig2} in Appendix~\ref{app_chandraimages} is an expansion of Fig.~\ref{fig_chandra} and shows the sources from Fig.\ref{fig_chandra} as well as the brightest source in each of our {\em Chandra} observations on top of the corresponding {\em Chandra} image. 

In summary, our astrometry indicates that the brightest X-ray source in the core of Terzan\,6 that is detected when \grs\, is in quiescence is {\em not} \grs. The light curve of X2 is distinctly different from that of \grs~as we demonstrate in the following section.

\subsubsection{Light curves and eclipses}\label{sec:res_eclipses}

\begin{figure*}[t]
\centerline{\includegraphics[width=18cm]{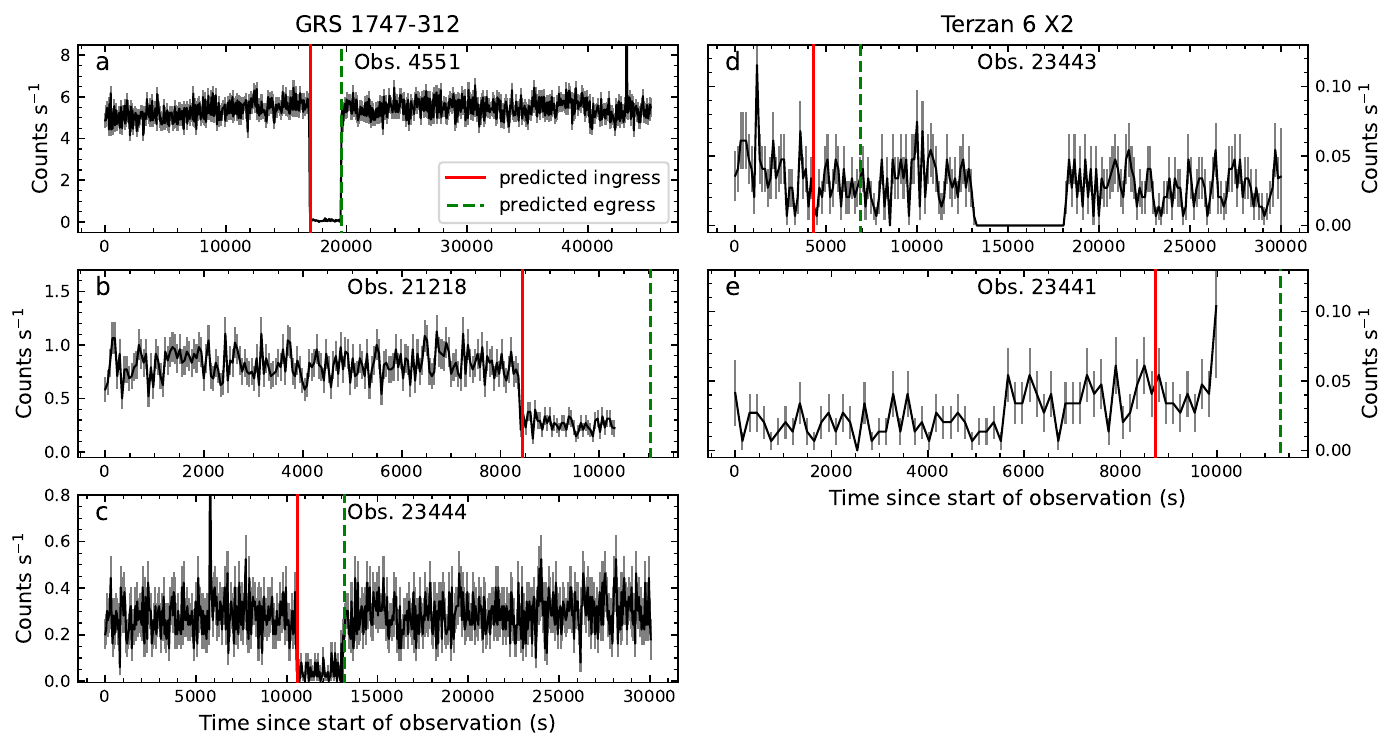}}
\caption{Barycentered light curves of the brightest source in five of the six {\em Chandra} observations. No eclipse from \grs\ was expected in ObsID 720 and none was observed. The left column shows light curves from when \grs\ was in outburst; these have a time resolution of 50\,s. The right column shows light curves of Terzan 6 X2, from when \grs\ was in quiescence; these have a time resolution of 150\,s. The solid red and dashed green lines show the expected ingress and egress times for \grs. Error bars are shown in gray.}
\label{fig:chandra_lc}
\end{figure*}

In Fig.\ \ref{fig:chandra_lc} we show the \chandra\ light curves of the brightest source in each of our observations. Light curves from observations made when \grs\ was confirmed to be in outburst are shown on the left; on the right we show light curves from observations made when \grs\ was in quiescence and X2 was the brightest source. Since we focus on the presence of eclipses, the light curve of observation 720 is not included; no eclipse was expected for \grs\ and none was observed.

As can be seen, the outburst light curves show eclipses at the predicted times for \grs\ (using the ephemeris from \citet{intzea03}): predicted eclipse ingresses and egresses are shown with solid red and dotted green lines, respectively. The two light curves of X2 do not show eclipses at the expected time for \grs. However, a clear eclipse is detected in observation 23443; it shows up as a $\sim$5 ks stretch in which no photons were detected in the extraction region. The length of the eclipse can be determined more accurately from a light curve with the native ACIS-S time resolution (3.14104\,s). We define the length of this eclipse as the time of the first detection at the end of the eclipse minus the last detection at the start of the eclipse. This yields a length of 5140\,s. This is substantially longer than the eclipses detected in the outburst light curves of \grs\ (2.6 ks; \citet{intzea03}). Note that this value should be considered an upper limit. The average time between photon arrivals is $\sim$30\,s, so a more likely eclipse length would be 5110\,s (as, on average, the eclipse will start, and end, halfway between photon arrival times). In addition to the eclipse, there are hints of intrinsic variability in the light curve of observation 23443. This is also the case for observation 23441.

In the right panel of Fig.\ \ref{fig_chandra} we show an image from the eclipse interval of observation 23443. X2 is not detected in this image, but a contribution from the faint nearby sources can clearly be seen. The position of \grs\ is marked as well. From the number of counts from \grs\ in the eclipse interval of X2 (5 in 5140\,s; 0.5\arcsec\ extraction radius), we estimate the luminosity of \grs\ in quiescence. Assuming an $N_{\rm H}$ of $1.9\times10^{22}$ cm$^{-2}$ and a power-law spectrum with index 1.7  (i.e., $\propto E^{-1.7}$), we find, using PIMMS\footnote{\url{https://asc.harvard.edu/toolkit/pimms.jsp}}, a luminosity of $\sim$1.7$\times10^{32}$ \lum\ (0.5--10 keV). Instead, When using the 3$\sigma$ upper limit on the number of counts \citep[16.03;][]{gehrels1986}, we find an upper limit on the luminosity of 5.4$\times10^{32}$ \lum.

\begin{figure*}[t]
\centerline{\includegraphics[width=18cm]{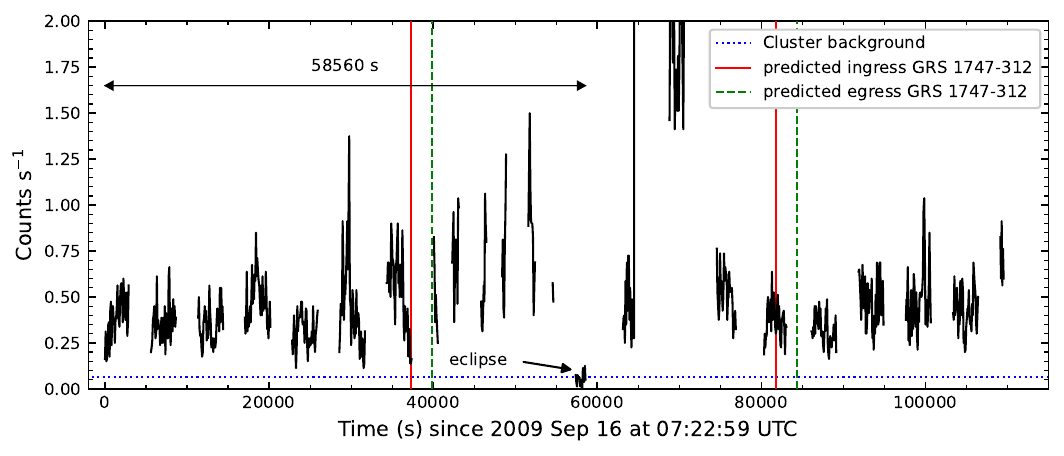}}
\caption{Barycentered {\em Suzaku} light curve of \ter. The energy band is 0.5--10 keV and the time resolution is 80\,s. A type-I X-ray burst started around t=64500 and peaked at a count rate of $\sim$113 counts\,s$^{-1}$. The y-scale was limited to a maximum value of 2 counts\,s$^{-1}$, to allow details at low count rates to remain visible. The expected eclipse ingress and egress times for \grs\ are shown with red solid and green dashed lines, respectively. A possible eclipse from \ter\ X2 around t=57500\,s is marked; count rates during this eclipse are consistent with the cluster background count rate (blue dotted line) as converted from {\it Chandra} data (see text). The long horizontal double-headed arrow indicates the lower limit on the orbital period of \ter\ X2.}
\label{fig:suzaku}
\end{figure*}

In Fig.\ \ref{fig:suzaku} we show the {\em Suzaku} 0.5--10 keV light curve of \ter. \grs\ was not in outburst during this observation. The sharp increase around $t=65$\,ks is due to a type-I X-ray burst. An analysis of this radius-expansion burst can be found in \citet{iwai2014}. Two eclipses from \grs\ were predicted to occur during the {\em Suzaku} observations. The expected ingress and egress times are shown as solid red and dashed green lines, respectively. The second ingress falls in the middle of a data segment and, as can be seen, no drop in the count rate is observed. The first ingress falls near the end of a data segment; only one data point is present after the predicted ingress, but it does not show a drop in count rate either (see also \cite{sajiea16}). The predicted egress times fall in between data segments. In addition to the missing eclipses in epochs 6 and 8, these constitute two additional cases of non-detections of eclipses when \grs\ is in quiescence. 

\citet{sajiea16} reported a strong drop in count rate around t=57500\,s; this drop is also clearly visible in our light curve. The count rate during the drop is consistent with the cluster background count rate when \grs\ is in quiescence and X2 in eclipse. This count rate was obtained in the following manner: we extracted a spectrum from a 15\arcsec-radius circle around  X2 during the eclipse segment in {\em Chandra} observation 23443 (epoch 6). This spectrum was fit with an absorbed power-law; the $N_{\rm H}$ was fixed to $1.9\times10^{22}$ cm$^{-2}$ (see Section \ref{sec:res_xspec}) and the power-law index was found to be 1.93. Using PIMMS this spectral shape was converted to a {\it Suzaku} count rate: 0.033 counts\,s$^{-1}$. This count rate was added to the normal {\em Suzaku} background in a 75\arcsec-radius circle (0.026 counts\,s$^{-1}$), resulting in an expected count rate during an eclipse of X2 of $\sim$0.06 counts\,s$^{-1}$ (XIS0, XIS1, and XIS3 combined). The fact that the count rate during the drop is consistent with the cluster background suggests that the drop might be another eclipse from \ter\ X2. No ingress or egress is observed for this eclipse, but the length of this data segment (1200\,s) does not exceed the eclipse duration measured for X2 with {\em Chandra}. We further note that the observed count rate during the eclipse of X2 puts a 3$\sigma$ upper limit on the 0.5--10 keV luminosity of \grs\ during the {\it Suzaku observation} of $\sim$4$\times10^{33}$ \lum\ (assuming an $N_{\rm H}$ of $1.9\times10^{22}$ cm$^{-2}$ and a power-law index of $1.7$).

Using the {\em Suzaku} light curve we can set a lower limit on the orbital period of X2.  None of the data gaps in the {\it Suzaku} data are longer than 5140\,s (the eclipse duration of X2 in epoch 6, so if other eclipses had occurred during the {\it Suzaku} observation an ingress and/or egress would have been detected. The best constraint on the orbital period can be determined from the interval between the start of the observation and the end of the eclipse data segment. The length of this interval is 58560\,s, and it is shown by the long horizontal double-headed arrow in the light curve. From this length a lower limit on the orbital period of 16.27 hr can be derived. 

Significant variability can be seen in the {\em Suzaku} light curve, with count rates fluctuating by a factor of $\sim$13 (ignoring the eclipse and the type-I X-ray burst). 

\subsection{X-ray spectra of X2} \label{sec:res_xspec}

We analysed the spectra of the two quiescent \chandra\ observations (23443 and 23441) and the eclipse interval from \chandra\ observation 23444. In all cases the flux is likely dominated by X2. The main goal of our spectral modelling is to obtain luminosity estimates, since even the spectrum with the most counts (1138 in observation 23443) does not allow for a detailed spectral decomposition. The three spectra were fit simultaneously. 

The first model we tried is an absorbed power law, using the \citet{wilms2000} X-ray absorption model: {\tt tbabs $\times$ pegpwrlaw}, where {\tt tbabs} is the absorption component and {\tt pegpwrlaw} is the power-law component ($\propto E^{-\Gamma}$). The $N_{\rm H}$ and power-law index ($\Gamma$) were linked between all three spectra, since we did not expect major variations in the interstellar absorption and the power-law slope between observations; the power-law normalizations were left to vary independently. This results in a very poor fit, with a reduced $\chi^2$ of 2.81 for 43 degrees of freedom (d.o.f.) and a null hypothesis probability $p_{\rm null}$ of 2.6$\times10^{-9}$. Leaving the power-law indices to vary independently improves the fit ($\chi^2_{\rm red}$/d.o.f.=1.47/41, $p_{\rm null}=2.7\times10^{-2}$), but for observation 23441 the power-law index $\Gamma$ takes on an unphysically low value ($-0.3\pm0.2$). Next, we allowed the $N_{\rm H}$ values to vary independently, but kept the power-law indices linked. This also improves on our initial fit, but not as much as the previous fit ($\chi^2_{\rm red}$/d.o.f.=1.81/41, $p_{\rm null}=1.1\times10^{-3}$). The $N_{\rm H}$ is found to vary between $\sim$$7.6\times10^{21}$ cm$^{-2}$ and $\sim$$5.7\times10^{22}$ cm$^{-2}$, while we obtain a power-law index of 0.96$\pm$0.15. 

The varying $N_{\rm H}$ values in the last fit hint at strong changes in the local absorption in the system. Such changes are often observed in eclipsing and dipping systems and are generally modeled by adding a partial covering absorption component to spectral models \citep{church1998}. We follow this practice by adding the XSPEC partial covering component {\tt pcfabs} to our model: {\tt tbabs $\times$ pcfabs $\times$ pegpwrlaw}. We link the values of the {\tt tbabs} $N_{\rm H}$ parameter and the power-law index and let the {\tt pcfabs} parameters vary independently. Only for observation 23441 does the $N_{\rm H}$ value of the {\tt pcfabs} component differ significantly from zero. For observation 23443 and the eclipse segment of 23444 these values were very small, meaning that the covering fraction (F$_{\rm cov}$) became meaningless. For those two spectra we fixed both {\tt pcfabs} parameters to zero. The resulting fit is still not satisfactory, with $\chi^2_{\rm red}$/d.o.f.=1.56/41 and $p_{\rm null}=1.3\times10^{-2}$.

Like for other eclipsing sources, the variability in the light curves of X2 (see Figs.\ \ref{fig:chandra_lc} and \ref{fig:suzaku}) may be the result of changes in the absorption column \citep{heinkeea03,wijnandsea03,diaztrigo06}. We therefore extract two spectra for  observation 23443 (which is the longest of the three): a high-count-rate spectrum ($\ge0.06$ counts\,s$^{-1}$, when using a light curve from a 1\arcsec-radius extraction region, instead of the 0.5\arcsec-radius extraction region used for the light curve in Fig.\ \ref{fig:chandra_lc}d) and a low-count-rate spectrum ($<0.06$ counts\,s$^{-1}$), which presumably has a higher absorption column. The eclipse interval was not included in the spectral extraction. The high- and low-count-rate spectra had 728 and 411 counts, respectively. In our simultaneous fit we replace the spectrum from observation 23443 with the high- and low-count-rate spectra from the same observation. The normalizations of the power-law component are linked for the latter two spectra, assuming that all the observed variability in the light curve is due to variations in the absorption. For the high-count-rate spectrum of observation 23443 and the spectrum of the eclipse segment of 23444, we again find very low absorption columns for the {\tt pcfabs} component and for those spectra we set both the absorption column and covering fraction of the {\tt pcfabs} component to zero. The resulting fit is acceptable, with $\chi^2_{\rm red}$/d.o.f.=1.22/57, and $p_{\rm null}=0.22$. The fit results are shown in Table \ref{tab:spectra}. For a distance of 6.7 kpc the measured 0.5--10 keV fluxes correspond to luminosities of $\sim$0.4--2.0$\times10^{34}$ \lum. For comparison, the 0.5--10 keV luminosity during the 2009 {\em Suzaku} observation was $\sim5.9\times10^{34}$ erg\,s$^{-1}$ \citep[taken from][but using a 6.7 kpc distance, instead of  the 9.5 kpc that they used]{sajiea16}.

\begin{table}[t]
\caption{Spectral fit results from fits with the model: {\tt tbabs $\times$ pcfabs $\times$ pegpwrlaw}. All errors are 1$\sigma$.}
\begin{center}
\begin{tabular}{ccccc}
\hline
\hline
Epoch  &  6 & 6 & 7$^a$ & 8 \\
ObsID  &  23443 & 23443 & 23444$^a$ & 23441 \\
  &  high rate & low rate &  &  \\

\hline
 $N_{\rm H}$$^b$       & \multicolumn{4}{c}{(1.9$\pm$0.3)} \\
 $\Gamma$     & \multicolumn{4}{c}{1.34$\pm$0.16}\\

 F$_{\rm cov}$     & 0$^c$ & 0.68$\pm$0.03 & 0$^c$ & 0.86$\pm$0.03 \\
 $N_{\rm H, cov}$$^b$& 0$^c$ & 160$^{+150}_{-50}$ & 0$^c$ & 20$\pm$7 \\
$F_X^d$ & \multicolumn{2}{c}{2.80$\pm$0.16} & 0.80$\pm$0.13 & 3.7$^{+1.0}_{-0.7}$ \\   
$L_X^e$ & \multicolumn{2}{c}{1.50$\pm$0.09} & 0.43$\pm$0.07 & 2.0$^{+0.5}_{-0.4}$ \\   \hline 
\end{tabular}
\end{center}
\noindent$^a$ Only the eclipse interval was used\\
$^b$ 10$^{22}$ cm$^{-2}$ \\
$^c$ parameter was fixed \\
$^d$ 0.5--10 keV unabsorbed flux (10$^{-12}$ erg\,s$^{-1}$\,cm$^{-2}$)   \\
$^e$ 0.5--10 keV luminosity (10$^{34}$ \lum)  
\label{tab:spectra}
\end{table}

\subsection{Search for optical counterparts} \label{sec:res_opt}

We have searched the {\em HST} images for optical counterparts to X2 and \grs. Multiple optical sources lie inside the 2$\sigma$ error radii.  To determine whether the true counterpart is among these candidates, we have examined the $V_{\rm 606}$ and $I_{\rm 814}$ magnitudes and colors, and the variability of these stars.

\begin{figure}
\centerline{\includegraphics[width=8.5cm]{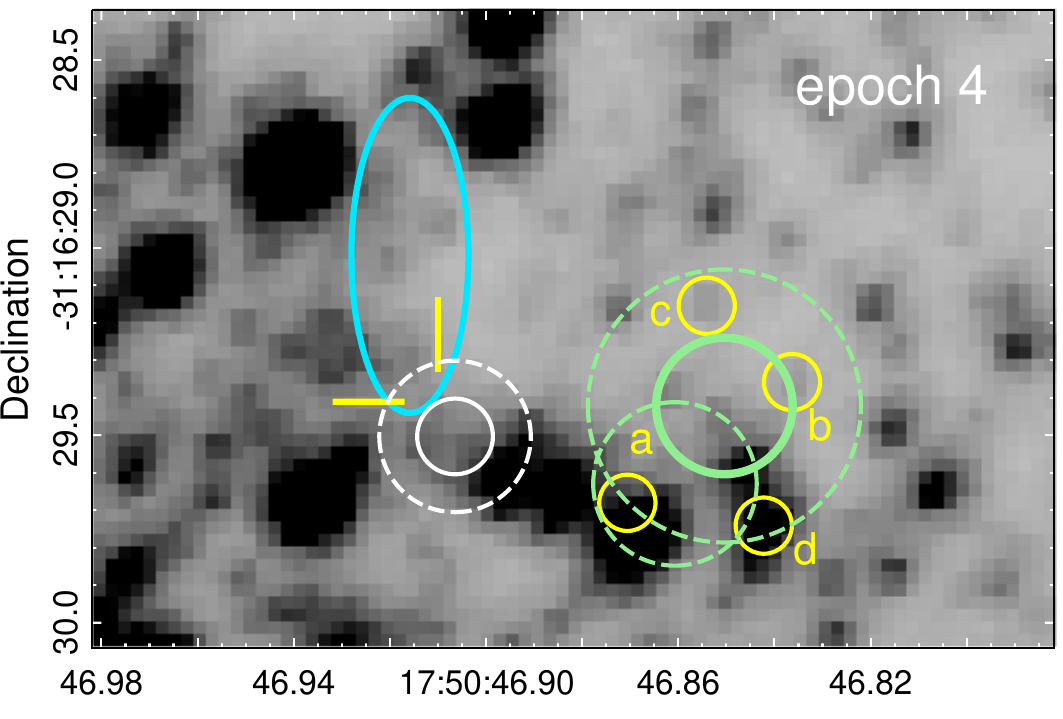}} 
\centerline{\includegraphics[width=8.5cm]{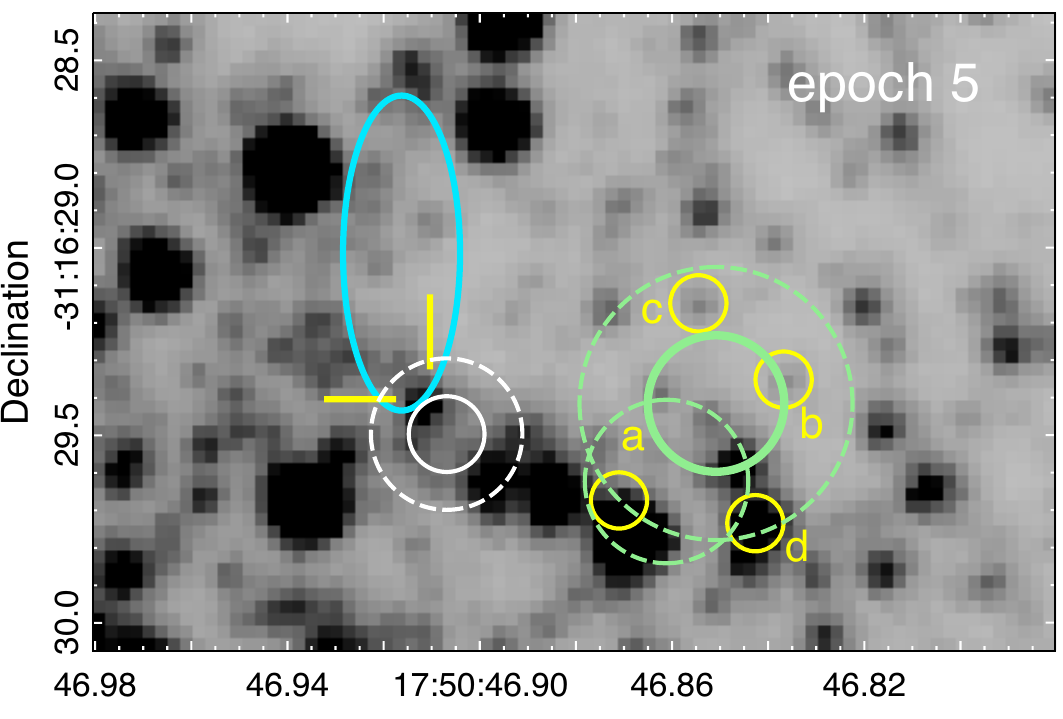}} 
\centerline{\includegraphics[width=8.5cm]{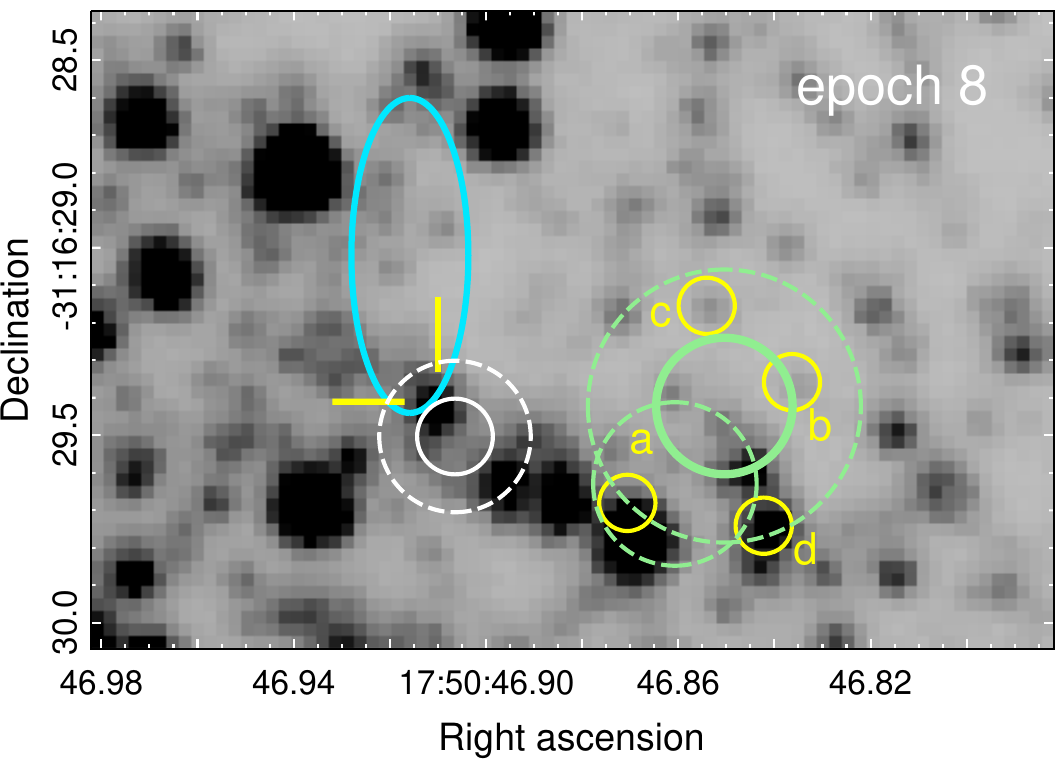}} 
\caption{F606W images from epoch 4 ({\em top}), epoch 5 ({\em middle}) and epoch 8 ({\em bottom}). The 1$\sigma$ and 2$\sigma$ error circles on the {\em Chandra} positions of X2 (white) and \grs\ (green) are marked with solid and dashed circles, respectively. The yellow tick marks indicate our suggested optical counterpart for X2. The $V_{\rm 606}$ magnitude difference between epochs 4 and 5 is $\sim$2.1, and between epochs 5 and 8 $\sim$0.6. Optical variables near \grs\ are marked with $a$, $b$, $c$ and $d$ and are discussed in the text. The blue ellipse shows the 1$\sigma$ radio error circle from the ATCA 5.5 GHz observation presented in \cite{panuea21}. North is up, east to the left. \label{fig_hst}}
\end{figure}

\begin{figure}
\includegraphics[width=8.5cm,bb=20 40 520 260]{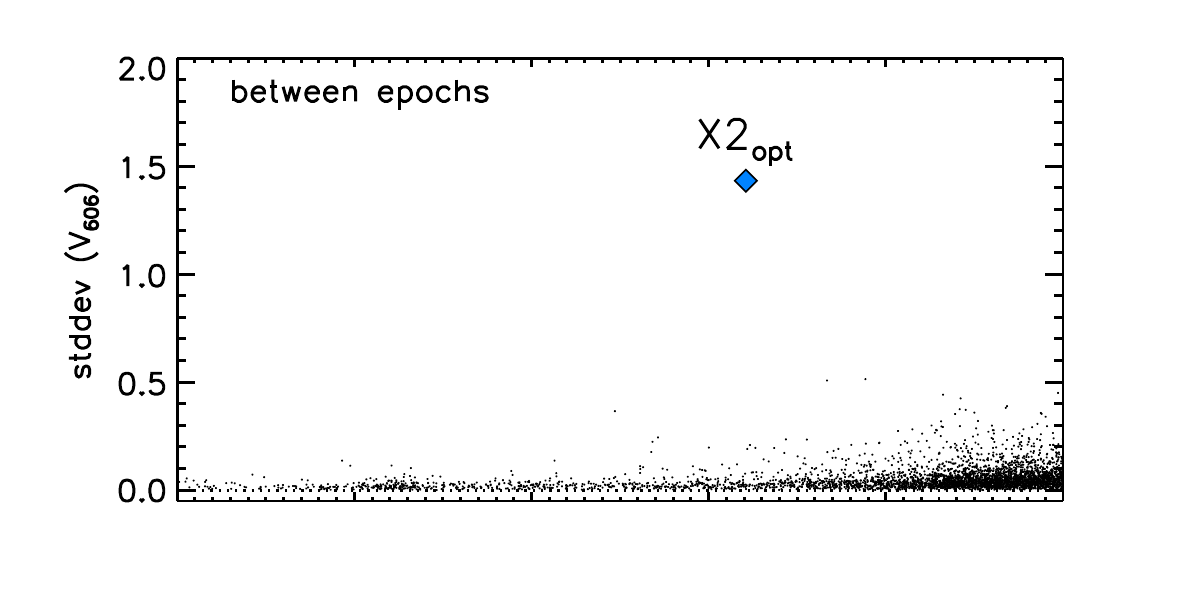}\\
\includegraphics[width=8.5cm,bb=20 20 520 275]{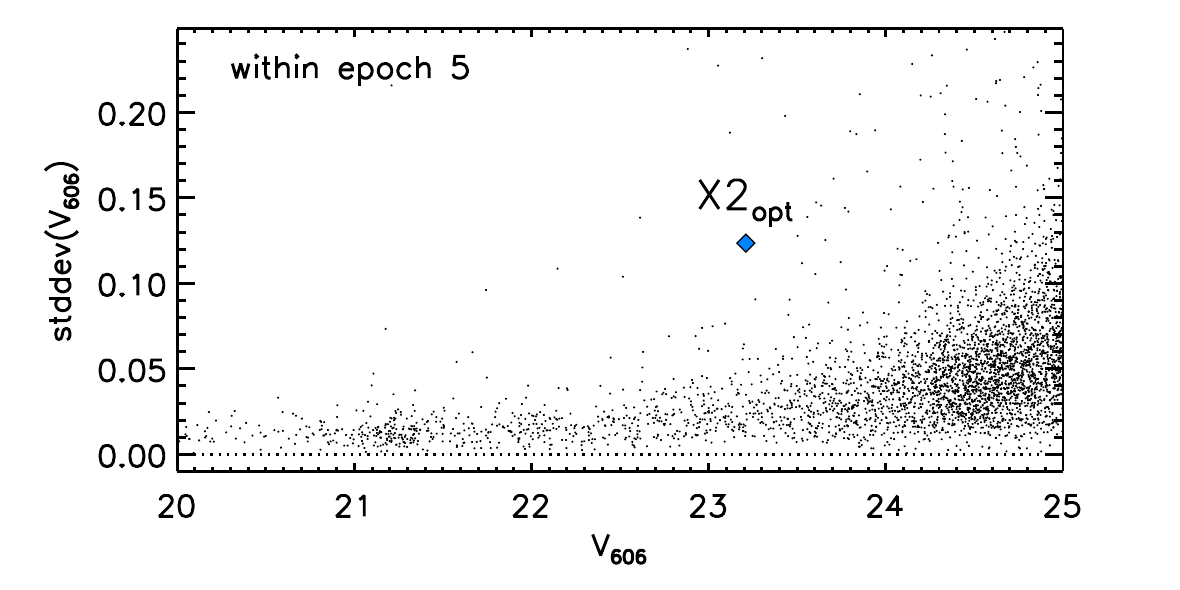}
\caption{{{\em Top:} Standard deviation in the average $V_{\rm 606}$ of epochs 4, 5 and 8 as function of $V_{\rm 606}$. \em Bottom:} Standard deviation in $V_{\rm 606}$  as measured from the individual long exposures of GO\,15616 as a function of $V_{\rm 606}$. Only stars inside the half-light radius of \ter\ are plotted. X2$_{\rm opt}$ is marked with a blue diamond. \label{fig_std}}
\end{figure}

\begin{figure}
\includegraphics[width=8.5cm]{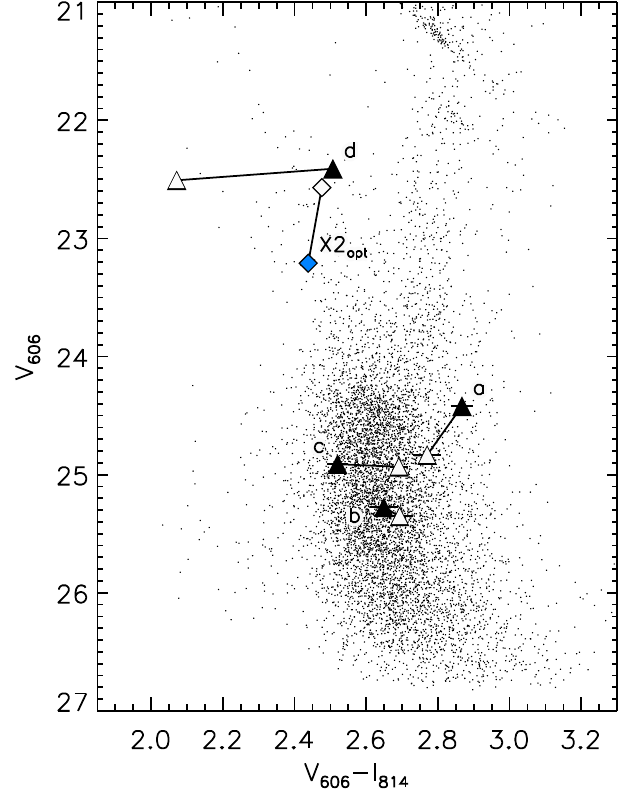}
\caption{GO\,15616 $V_{{\rm 606}}$ versus $V_{{\rm 606}}-I_{{\rm
      814}}$ color-magnitude diagram of \ter\, showing stars inside the half-light radius. Photometry for the stars discussed in the text is shown with filled symbols for GO\,15616 (epoch 5) and open symbols for GO\,16420 (epoch 8). The plausible counterpart for X2 is plotted with a diamond. Stars $a$, $b$, $c$ and $d$  in the 2$\sigma$ error circle of \grs\ are plotted with triangles. \label{fig_cmd}}
\end{figure}

We analyzed {\em HST} imaging from three epochs: epoch 4 ($V_{\rm 606}$) and epochs 5 and 8 ($V_{\rm 606}$, $I_{\rm 814}$). For X2, we estimate that the X-ray luminosity in epoch 8 is $L_{\rm X} \approx 2 \times 10^{34}$ erg s$^{-1}$ (0.5--10 keV) based on the X-ray spectral fits from Section \ref{sec:res_xspec}. We have no estimate of the X-ray luminosity in epochs 4 and 5, and therefore no a priori expectation for the optical variation of X2 between epochs. However, in the 2$\sigma$ error circle of X2, there is one star that stands out from the others given its large-amplitude variations between epochs 4, 5 and 8. We have indicated this star (X2$_{\rm opt}$) with yellow tick marks in Fig.~\ref{fig_hst}. In epoch 5 this star brightened by 2.1 mag in $V_{\rm 606}$ compared to epoch 4, going from $V_{\rm 606}=25.3$ to $23.2$ (see the top and middle panels of Fig.~\ref{fig_hst}). In epoch 8, the star was even brighter, $V_{\rm 606}=22.6$ (bottom panel of Fig.~\ref{fig_hst}). In $I_{\rm 814}$ it brightened from $I_{\rm 814}=20.8$ to $I_{\rm 814}=20.1$ between epoch 5 and 8. Such large-amplitude variability is rare among the stars inside the half-light radius of Terzan\,6 as can be seen in Fig.~\ref{fig_std} (top), which shows the standard deviation of the $V_{\rm 606}$ magnitudes between the three {\em HST} epochs as a function of magnitude. The photometry of X2$_{\rm opt}$ places it 1.1--1.7 mag above the main-sequence turnoff in the color-magnitude diagram (CMD) of \ter~(see Fig.~\ref{fig_cmd}). No F814W images were taken during GO\,14074 so we do not know how it changed color as the star drastically brightened in $V_{\rm 606}$. Compared to other stars, X2$_{\rm opt}$ also shows enhanced variability between the individual exposures, in particular in $V_{\rm 606}$ (Fig.~\ref{fig_std}, bottom). The light curve of the star shows variability in epoch 5 with peak-to-peak variations of 0.3 mag in $V_{\rm 606}$ and 0.16 mag in $I_{\rm 814}$ between the four long exposures per filter. We consider star X2$_{\rm opt}$ the plausible optical counterpart of X2. The angular separation between X2 and X2$_{\rm opt}$ is 0.10\arcsec.

 We know that in X-rays \grs~was in quiescence in epoch 4 (it had returned to quiescence more than a month before), in outburst in epoch 5, and again in quiescence in epoch 8 (when it had returned to quiescence about two months before). Therefore, we expect that the true optical counterpart to \grs~ can be identified by looking for stars that are brighter in epoch 5 compared to epochs 4 and 8\footnote{None of the {\em HST} exposures were taken around phase 0 (i.e.~the phase of the X-ray eclipse, see \cite{intzea03}). Therefore, we do not expect that the optical magnitudes are temporarily dimmed by an eclipse of portions of the disk by the secondary.} (e.g. \citealt{pallea13,ferrea15}). There is only one star (marked $a$ in Fig.~\ref{fig_hst}) that satisfies this condition in the area where the 2$\sigma$ error circles of ObsID 720 and ObsID 21218 overlap. In epoch 5, star $a$ was 0.2 and 0.4 mag brighter in $V_{\rm 606}$ than in epochs 4 and 8, respectively; in epoch 5 this star was brighter by 0.5 mag in $I_{\rm 814}$ compared to epoch 8. However, since this is a faint star ($V_{\rm 606} \approx 24.4$) that lies in the wings of a much ($>$3 mag in $V_{\rm 606}$) brighter star, we do not consider its photometry very reliable. Moreover, this star became bluer in epoch 8. This is not what is expected for an X-ray binary returning to quiescence. As the contribution from an accretion disk to the optical light declines as an LMXB enters quiescence, the colors of the true counterpart should become redder (if the mass donor is a non-degenerate star like in \grs, given its 12.4 hr orbital period). We have marked two more stars in Fig.~\ref{fig_hst}, $b$ and $c$, which lie in the 2$\sigma$ error circle of ObsID 720 but outside the 2$\sigma$ error circle of ObsID 21218. Both became brighter by $\sim$0.3 mag in $V_{\rm 606}$ going from epoch 4 to 5. However, star $b$ did not change brightness between epochs 5 and 8 (the change in $V_{\rm 606}$ and $I_{\rm 814}$ was $<$0.1 mag and not significant). The $V_{\rm 606}$ magnitude of star $c$ also remained constant, but it became slightly (0.14 mag) brighter in $I_{\rm 814}$ and as a result got redder as \grs\ went in quiescence. We do not consider stars $b$ and $c$ plausible counterparts to \grs: they lie relatively far from our most precise position for the source (from ObsID 21218) and the photometry in outburst is not (much) brighter than in quiescence. Stars $a$ to $c$ lie around, or $\sim$1 mag below, the main-sequence turnoff in the CMD of GO\,15616 and GO\,16420 (Fig.~\ref{fig_cmd}). The largest variability between epochs is seen in star $d$: in epoch 5, it was 0.53 mag brighter in $I_{\rm 814}$ with respect to epoch 8. However, it remained constant in $V_{\rm 606}$ to within 0.1 mag between all three epochs, and consequently star $d$ was much bluer ($V_{\rm 606}-I_{\rm 814}$ changed from 2.5 to 2.05) in the quiescent state of \grs. Like for the other stars, we do not consider $d$ a plausible counterpart.

\section{Discussion}\label{sec_dis}

We have performed \chandra\ observations of the globular cluster \ter\ when the recurrent transient \grs\ was in quiescence. These observations allowed us, for the first time, to explore the population of faint X-ray sources near the cluster's center. We found that, when \grs\ is in quiescence (ObsIDs 23443 and 23441), the cluster's X-ray emission is dominated by a source, \ter\ X2, that is spatially offset by 0.7\arcsec~from a \chandra\ position of \grs\ obtained during outburst (ObsID 720). This offset is 3.4 times larger than the error on the separation between the two sources after aligning the astrometry of the {\em Chandra} observations to that of the {\em HST} images. When \grs\ was in outburst during ObsID 720, eclipses at times expected for its accurate ephemeris were observed with other missions, leaving no doubt about the identity of the bright source in Terzan\,6. On the other hand, the absence of an eclipse at the predicted time for \grs\, and the observation of a 5140\,s-long eclipse (nearly twice as long as the eclipses in \grs) at a later time instead, clearly demonstrate that the source detected in ObsID 23443 is {\em not} \grs. Our observation of ``missing" and ``unexpected" eclipses in the light curve of \grs\, is not the first instance when these have been reported. A strong eclipse candidate, as well as two ``missed" eclipses, were observed in a \suzaku\ observation that was made when \grs\ was in quiescence \citep{sajiea16}. From this \suzaku\ observation a lower limit on the orbital period of X2 can be determined: 16.27 hr.

Constraints on how the X-ray luminosity of X2 varies can be derived from our own findings and from the literature. The 0.5--10 keV luminosities of X2 in the \chandra\ observations range from $\sim$4$\times10^{33}$ \lum\ to $\sim$2$\times10^{34}$ \lum. The 0.5--10 keV luminosity during the 2009 {\em Suzaku} observation of \ter\ was $\sim$5.9$\times10^{34}$ erg\,s$^{-1}$, slightly above the range observed with \chandra. Combined with the absence of predicted eclipse ingresses for \grs, this suggests that the flux during the \suzaku\ observation was also dominated by X2. The burst observed during the \suzaku\ observation reached a luminosity $\sim$1.4$\times10^{38}$ \lum\ and showed a phase of strong radius expansion \citep{iwai2014}. The properties of this burst are very similar to those of a radius expansion burst detected in the direction of \ter\ with {\em RXTE} \citep{intzea03b}, when \grs\ was not in outburst. Although this burst was originally attributed to \grs, it is likely that it originated from X2 as well. {\em XMM-Newton} observed \ter\ in 2004, when \grs\ was in quiescence. \citet{vatsea18b} measured a 0.5--10 keV luminosity of $\sim$2.4$\times10^{33}$ \lum\ (corrected to a distance of 6.7 kpc). This is slightly below  the lowest luminosity observed for X2 in the \chandra\ observations. Combined with the very low quiescent luminosity of \grs\ (less than a few times 10$^{32}$ \lum), this suggests that X2 was dominating the flux from \ter\ in the {\em XMM-Newton} observation as well. The X-ray luminosity range of X2 overlaps with that of both the quiescent neutron-star LMXBs \citep[$<5\times10^{33}$ \lum;][]{armas2014} and the persistent very-faint X-ray binaries ($10^{34}$--$10^{36}$ \lum; see \citet{bahrdege23} for an overview). Due to the presence of a type I X-ray burst in the {\it Suzaku} observation, the compact object in X2 can be identified as a neutron star, like most of the persistent very-faint X-ray binaries.  We note that the quiescent luminosity of \grs\ (a few times $10^{32}$ \lum; see Section \ref{sec:res_eclipses}) is consistent with that of other quiescent neutron-star LMXBs in the same period range  \citep[$\sim$10--15 h;][]{armas2014}.

In the optical, the counterpart of X2 is $V_{\rm 606}$=25.3 at its faintest (in epoch 4). We have no information on the optical color and the X-ray state of X2 at this time, and consequently it remains unclear whether emission from an accretion disk contributes to the optical light. However, given the lower limit on the orbital period from {\em Suzaku}, we know that a Roche-lobe filling mass donor cannot be a small star (e.g.~$R_2\approx0.87$ $R_{\odot}$ for a 1.4 $M_{\odot}$ neutron star and an $M_2$=0.2 $M_{\odot}$ donor in a 16.27 hr orbit, see \cite{eggl83}) and must therefore contribute significantly to the optical emission in epoch 4\footnote{From comparison with stellar models, it follows that such a star is significantly bloated with respect to an undisturbed 0.2 $M_{\odot}$ star, which has $R\approx0.22$ $R_{\odot}$ \citep{bresea12}.}. If the orbital period of X2 can be further constrained, or measured, with follow-up observations, the ratio of the eclipse duration and the orbital period can set more stringent limits on the mass ratio $q$ and inclination of X2. From the current upper limit on this ratio (5140 s / (16.27$\times$3600 s) $\approx$ 0.088), it follows that $q\gtrsim0.31$ or $M_2\gtrsim0.43$ $M_{\odot}$. For larger periods, the lower limit on $q$, and hence on $M_2$, decreases (see the curves in Fig.~2 in \citealt{horn85}). For the upper limit on $M_2$, one can assume the donor must be less massive than the turn-off mass of Terzan\,6 ($\sim$0.8 $M_{\odot}$).

\ter\ has also been studied in the radio, when \grs\ was in quiescence. \citet{panuea21} present an analysis of near-simultaneous 2015 and 2018 radio (VLA and ATCA) and X-ray ({\it Swift}/XRT) observations. A single radio source was detected near the cluster core, located near our position of X2 (see Fig.~\ref{fig_hst}). The 5 and 5.5 GHz flux densities of the source ranged from $<$12.9 $\mu$Jy to more than 210 $\mu$Jy. Although {\it Swift}/XRT cannot resolve the X-ray population of \ter, we note that the X-ray luminosity range of the source in both 2015 and 2018, as reported by \citet{panuea21}, is comparable to that seen with {\it Chandra} for X2, $\sim$(0.7--4)$\times10^{34}$ \lum.   Furthermore, these X-ray luminosities are much higher than the quiescent X-ray luminosity of \grs, suggesting that it is X2 that is detected in the radio and X-ray, and not \grs. Indeed, \citet{panuea21} considered the possibility that this radio emission was from a second LMXB (which had to be highly radio-variable, X-ray bright at the $10^{34}$ \lum\ level in both 2015 and 2018, and only 0.7\arcsec\ from \grs), but decided it was more parsimonious to assume an astrometric error. We note that in the radio luminosity vs.\ X-ray luminosity diagram of \citet{panuea21} X2 is the radio brightest neutron-star LMXB in its X-ray luminosity range, on par with black-hole LXMBs. The only other neutron-star LMXBs detected in the radio in this X-ray luminosity range are transitional millisecond pulsars and accreting millisecond X-ray pulsars.

\begin{figure}
\includegraphics[width=8.5cm]{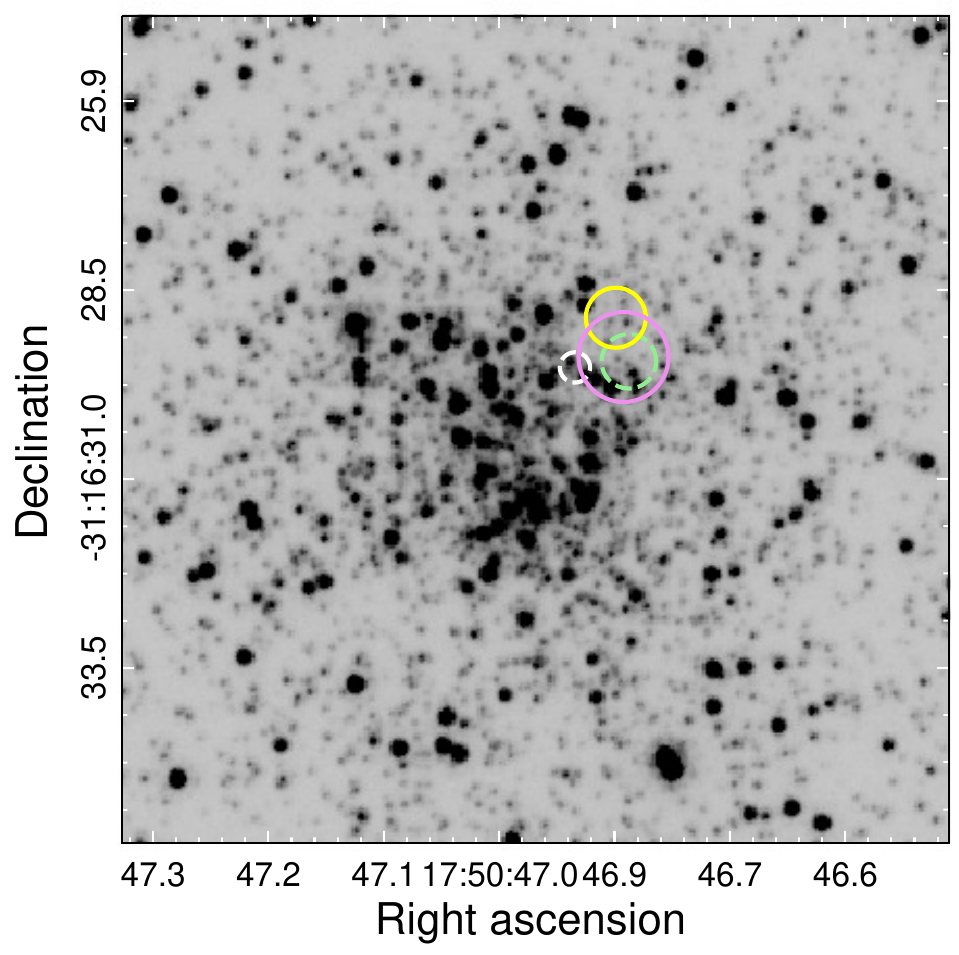}
\caption{Stacked F606W image from epoch 5, with the 2$\sigma$ error circles on the positions of \grs\ and X2 marked in green and white, respectively. The yellow circle shows the position of \grs\ from \cite{intzea03} (radius is their 95\% positional uncertainty of 0.4\arcsec) and the purple circle is their position of the cluster center (0.6\arcsec\,uncertainty).\label{fig_center}  }
\end{figure}

\cite{intzea03} redetermined the center of \ter\ based on iteratively calculating the mean position of stars within 15\arcsec\ of an initially chosen center. Their star catalogs were based on optical images obtained with the New Technology Telescope. While these images were obtained under excellent seeing conditions (0.6\arcsec), the resolution of our {\em HST} images is better by more than an order of magnitude. Fig.~\ref{fig_center} shows the center of \ter\ from \cite{intzea03} on top of our stacked F606W image from epoch 5. Clearly, their center does not align with the highest concentration of resolved stars, which appears to be offset to the south-east. A new determination of the cluster center is outside the scope of this paper. However, we do note that the highest concentration of stars in the {\em HST} images also does not seem to align with the peak of the X-ray emission in the {\em Chandra} images. Remeasuring the X-ray and optical center of the cluster could shed more light on the dynamical history of \grs, X2 and other X-ray sources, and whether an explanation is needed to account for a possible offset away from the optical cluster center. We defer such a discussion to a follow-up study as it requires a full consideration of the X-ray sources in \ter, including the fainter ones that we have detected.

Recently, \cite{painter2023} offered alternative interpretations of the eclipse behavior in the \chandra\ observations. They believe that the offset between the outburst position of \grs\ and the position of the source detected when \grs\ is not in outburst is within \chandra's astrometric uncertainty, while we believe that the offset is real and significant at the 3.4$\sigma$ level. \cite{painter2023} consider it unlikely that two independent (unrelated) eclipsing X-ray binaries are found on the edge of the cluster center at very close proximity ($\lesssim$0.7\arcsec) from each other.  A more likely scenario, according to \cite{painter2023}, is that \grs\ is part of a physically bound system of higher-order (triple or more) multiplicity. For more details we refer the reader to the paper by \cite{painter2023}. However, regarding the probability of finding two closely separated eclipsing neutron-star X-ray binaries, we first note that the concentration of X-ray binaries in globular clusters is enhanced to begin with, and it is not unprecedented to find two X-ray binaries at a small ($\sim$1\arcsec) projected distance (see the case of the two X-ray binaries in Liller 1, \citealt{homaea18}). Second, 12 of the $\sim$130 neutron-star LMXBs are eclipsing \citep{bahrdege23}; if one member of a pair of two closely separated neutron-star LMXBs is eclipsing, there is a $\sim$10\% chance that the other one shows eclipses as well. Finally, as we argue above, it is not exactly clear how far offset \grs\ and X2 are from the cluster center, and whether their location in the cluster puts constraints on a (joint) dynamical formation history.

In Fig.~\ref{fig_center} we show the position of \grs\ from \cite{intzea03} in yellow. Their position is 0.6\arcsec~different from the position that we measured from the same {\em Chandra} image (Table~\ref{tab_xpos}) but their 95\% uncertainty radius and our 2$\sigma$ error circle overlap. Differences in the boresighting (they used the USNO-B1.0 catalog, we used Gaia DR2; the X-ray sources that they used are not the same as the ones we used) can explain the small positional offset.

Despite having obtained {\em HST} imaging of \ter\ during outburst and quiescence of \grs\, we have not been able to unambiguously identify a plausible optical counterpart. We have discussed several stars near the {\em Chandra} position of \grs\ that were brighter during outburst, but the observed optical variations are modest ($<$0.6 mag). Reprocessing of X-rays in the accretion disk gives rise to a relation between the X-ray luminosity, optical luminosity (or absolute magnitude) and the orbital period (a measure of the size of the accretion disk) of X-ray binaries in outburst \citep{vanpmccl94}. Since we know the X-ray luminosity ($3.8\times10^{38}$ \lum\ at outburst peak for a distance of 6.7 kpc) and orbital period (12.3595 hr) of \grs\ \citep{intzea03}, we can use this relation to estimate the expected observed $V$ magnitude of the optical counterpart: $M_V\approx23.1\pm0.5$, where the range represents the scatter around this empirical relation. The variable stars $a$, $b$ and $c$ are all fainter (Fig.~\ref{fig_cmd}). However, \grs\ is seen at high inclination, and as a result the X-ray and optical emission from the system along our line of sight may be much reduced. 

\begin{acknowledgments}

This research has made use of data obtained from the Chandra Data Archive and the Chandra Source Catalog, and software provided by the Chandra X-ray Center in the application package CIAO. Support for this work was provided by the National Aeronautics and Space Administration through Chandra Award Numbers GO9-20033X, GO1-22041X, and GO1-22042X, issued by the Chandra X-ray Center, which is operated by the Smithsonian Astrophysical Observatory for and on behalf of the National Aeronautics Space Administration under contract NAS8-03060.

This research is based on observations made with the NASA/ESA Hubble Space Telescope obtained from the Space Telescope Science Institute, which is operated by the Association of Universities for Research in Astronomy, Inc., under NASA contract NAS 5–26555. These observations are associated with programs GO 15616 and GO 16420. Support for these programs was provided by NASA through a grant from the Space Telescope Science Institute, which is operated by the Association of Universities for Research in Astronomy, Inc., under NASA contract NAS 5-03127.

This research has made use of data and software provided by the High Energy Astrophysics Science Archive Research Center (HEASARC), which is a service of the Astrophysics Science Division at NASA/GSFC.

The authors would like to thank J.\,Strader for sharing their position of a radio source in Terzan 6 in an early phase of the manuscript.
 
\end{acknowledgments}

\facilities{CXO, HST, Suzaku, HEASARC, MAST}

\software{Astropy [ascl:1304.002], CIAO [ascl:1311.006], DOLPHOT [ascl:1608.013], DrizzlePac [ascl:1212.011], DS9 [ascl:0003.002], FTOOLS [ascl:9912.002], HEASOFT [ascl:1408.004], XSPEC [ascl:9910.005]}

\vspace{1cm}
{\em Note added in proof.} After acceptance of this manuscript we have been able to determine the orbital period of Terzan\,6 X2, using more recent observations of X-ray eclipses of the source. The orbital period is 21.17914(3) hr. We obtained the following ephemeris for mid-eclipse (in Barycentric Dynamical Time): $T_{mid}(n)$ = MJD 60195.14768(11) + 0.88246416(12)$n$.


\begin{thebibliography}{}
\expandafter\ifx\csname natexlab\endcsname\relax\def\natexlab#1{#1}\fi
\providecommand{\url}[1]{\href{#1}{#1}}
\providecommand{\dodoi}[1]{doi:~\href{http://doi.org/#1}{\nolinkurl{#1}}}
\providecommand{\doeprint}[1]{\href{http://ascl.net/#1}{\nolinkurl{http://ascl.net/#1}}}
\providecommand{\doarXiv}[1]{\href{https://arxiv.org/abs/#1}{\nolinkurl{https://arxiv.org/abs/#1}}}

\bibitem[{{Armas Padilla} {et~al.}(2014){Armas Padilla}, {Wijnands},
  {Altamirano}, {M{\'e}ndez}, {Miller}, \& {Degenaar}}]{armas2014}
{Armas Padilla}, M., {Wijnands}, R., {Altamirano}, D., {et~al.} 2014, \mnras,
  439, 3908, \dodoi{10.1093/mnras/stu243}

\bibitem[{{Arnaud}(1996)}]{arnaud1996}
{Arnaud}, K.~A. 1996, in Astronomical Society of the Pacific Conference Series,
  Vol. 101, Astronomical Data Analysis Software and Systems V, ed. G.~H.
  {Jacoby} \& J.~{Barnes}, 17

\bibitem[{{Bahramian} \& {Degenaar}(2023)}]{bahrdege23}
{Bahramian}, A., \& {Degenaar}, N. 2023, in Handbook of X-ray and Gamma-ray
  Astrophysics. Edited by Cosimo Bambi and Andrea Santangelo, 120,
  \dodoi{10.1007/978-981-16-4544-0_94-1}

\bibitem[{{Bahramian} {et~al.}(2013){Bahramian}, {Heinke}, {Sivakoff}, \&
  {Gladstone}}]{bahrea13}
{Bahramian}, A., {Heinke}, C.~O., {Sivakoff}, G.~R., \& {Gladstone}, J.~C.
  2013, \apj, 766, 136, \dodoi{10.1088/0004-637X/766/2/136}

\bibitem[{{Bohlin}(2012)}]{bohl12}
{Bohlin}, R.~C. 2012, {Flux Calibration of the ACS CCD Cameras IV. Absolute
  Fluxes}, Instrument Science Report ACS 2012-01, 28 pages

\bibitem[{{Bressan} {et~al.}(2012){Bressan}, {Marigo}, {Girardi}, {Salasnich},
  {Dal Cero}, {Rubele}, \& {Nanni}}]{bresea12}
{Bressan}, A., {Marigo}, P., {Girardi}, L., {et~al.} 2012, \mnras, 427, 127,
  \dodoi{10.1111/j.1365-2966.2012.21948.x}

\bibitem[{{Church} {et~al.}(1998){Church}, {Ba{\l}uci{\'n}ska-Church},
  {Dotani}, \& {Asai}}]{church1998}
{Church}, M.~J., {Ba{\l}uci{\'n}ska-Church}, M., {Dotani}, T., \& {Asai}, K.
  1998, \apj, 504, 516, \dodoi{10.1086/306049}

\bibitem[{{Cohen} {et~al.}(2018){Cohen}, {Mauro}, {Alonso-Garc{\'\i}a},
  {Hempel}, {Sarajedini}, {Ordo{\~n}ez}, {Geisler}, \& {Kalirai}}]{coheea18}
{Cohen}, R.~E., {Mauro}, F., {Alonso-Garc{\'\i}a}, J., {et~al.} 2018, \aj, 156,
  41, \dodoi{10.3847/1538-3881/aac889}

\bibitem[{{D{\'\i}az Trigo} {et~al.}(2006){D{\'\i}az Trigo}, {Parmar},
  {Boirin}, {M{\'e}ndez}, \& {Kaastra}}]{diaztrigo06}
{D{\'\i}az Trigo}, M., {Parmar}, A.~N., {Boirin}, L., {M{\'e}ndez}, M., \&
  {Kaastra}, J.~S. 2006, \aap, 445, 179, \dodoi{10.1051/0004-6361:20053586}

\bibitem[{{Dolphin}(2000)}]{dolp00}
{Dolphin}, A.~E. 2000, \pasp, 112, 1383

\bibitem[{{Eggleton}(1983)}]{eggl83}
{Eggleton}, P.~P. 1983, \apj, 268, 368, \dodoi{10.1086/160960}

\bibitem[{{Ferraro} {et~al.}(2015){Ferraro}, {Pallanca}, {Lanzoni}, {Cadelano},
  {Massari}, {Dalessandro}, \& {Mucciarelli}}]{ferrea15}
{Ferraro}, F.~R., {Pallanca}, C., {Lanzoni}, B., {et~al.} 2015, \apjl, 807, L1,
  \dodoi{10.1088/2041-8205/807/1/L1}

\bibitem[{{Gaia Collaboration} {et~al.}(2018){Gaia Collaboration}, {Brown},
  {Vallenari}, {Prusti}, {de Bruijne}, {Babusiaux}, {Bailer-Jones}, {Biermann},
  {Evans}, {Eyer}, {Jansen}, {Jordi}, {Klioner}, {Lammers}, {Lindegren},
  {Luri}, {Mignard}, {Panem}, {Pourbaix}, {Randich}, {Sartoretti}, {Siddiqui},
  {Soubiran}, {van Leeuwen}, {Walton}, {Arenou}, {Bastian}, {Cropper},
  {Drimmel}, {Katz}, {Lattanzi}, {Bakker}, {Cacciari}, {Casta{\~n}eda},
  {Chaoul}, {Cheek}, {De Angeli}, {Fabricius}, {Guerra}, {Holl}, {Masana},
  {Messineo}, {Mowlavi}, {Nienartowicz}, {Panuzzo}, {Portell}, {Riello},
  {Seabroke}, {Tanga}, {Th{\'e}venin}, {Gracia-Abril}, {Comoretto},
  {Garcia-Reinaldos}, {Teyssier}, {Altmann}, {Andrae}, {Audard},
  {Bellas-Velidis}, {Benson}, {Berthier}, {Blomme}, {Burgess}, {Busso},
  {Carry}, {Cellino}, {Clementini}, {Clotet}, {Creevey}, {Davidson}, {De
  Ridder}, {Delchambre}, {Dell'Oro}, {Ducourant},
  {Fern{\'a}ndez-Hern{\'a}ndez}, {Fouesneau}, {Fr{\'e}mat}, {Galluccio},
  {Garc{\'\i}a-Torres}, {Gonz{\'a}lez-N{\'u}{\~n}ez}, {Gonz{\'a}lez-Vidal},
  {Gosset}, {Guy}, {Halbwachs}, {Hambly}, {Harrison}, {Hern{\'a}ndez},
  {Hestroffer}, {Hodgkin}, {Hutton}, {Jasniewicz}, {Jean-Antoine-Piccolo},
  {Jordan}, {Korn}, {Krone-Martins}, {Lanzafame}, {Lebzelter}, {L{\"o}ffler},
  {Manteiga}, {Marrese}, {Mart{\'\i}n-Fleitas}, {Moitinho}, {Mora}, {Muinonen},
  {Osinde}, {Pancino}, {Pauwels}, {Petit}, {Recio-Blanco}, {Richards},
  {Rimoldini}, {Robin}, {Sarro}, {Siopis}, {Smith}, {Sozzetti}, {S{\"u}veges},
  {Torra}, {van Reeven}, {Abbas}, {Abreu Aramburu}, {Accart}, {Aerts},
  {Altavilla}, {{\'A}lvarez}, {Alvarez}, {Alves}, {Anderson}, {Andrei},
  {Anglada Varela}, {Antiche}, {Antoja}, {Arcay}, {Astraatmadja}, {Bach},
  {Baker}, {Balaguer-N{\'u}{\~n}ez}, {Balm}, {Barache}, {Barata}, {Barbato},
  {Barblan}, {Barklem}, {Barrado}, {Barros}, {Barstow}, {Bartholom{\'e}
  Mu{\~n}oz}, {Bassilana}, {Becciani}, {Bellazzini}, {Berihuete}, {Bertone},
  {Bianchi}, {Bienaym{\'e}}, {Blanco-Cuaresma}, {Boch}, {Boeche}, {Bombrun},
  {Borrachero}, {Bossini}, {Bouquillon}, {Bourda}, {Bragaglia}, {Bramante},
  {Breddels}, {Bressan}, {Brouillet}, {Br{\"u}semeister}, {Brugaletta},
  {Bucciarelli}, {Burlacu}, {Busonero}, {Butkevich}, {Buzzi}, {Caffau},
  {Cancelliere}, {Cannizzaro}, {Cantat-Gaudin}, {Carballo}, {Carlucci},
  {Carrasco}, {Casamiquela}, {Castellani}, {Castro-Ginard}, {Charlot},
  {Chemin}, {Chiavassa}, {Cocozza}, {Costigan}, {Cowell}, {Crifo}, {Crosta},
  {Crowley}, {Cuypers}, {Dafonte}, {Damerdji}, {Dapergolas}, {David}, {David},
  {de Laverny}, {De Luise}, {De March}, {de Martino}, {de Souza}, {de Torres},
  {Debosscher}, {del Pozo}, {Delbo}, {Delgado}, {Delgado}, {Di Matteo},
  {Diakite}, {Diener}, {Distefano}, {Dolding}, {Drazinos}, {Dur{\'a}n},
  {Edvardsson}, {Enke}, {Eriksson}, {Esquej}, {Eynard Bontemps}, {Fabre},
  {Fabrizio}, {Faigler}, {Falc{\~a}o}, {Farr{\`a}s Casas}, {Federici},
  {Fedorets}, {Fernique}, {Figueras}, {Filippi}, {Findeisen}, {Fonti},
  {Fraile}, {Fraser}, {Fr{\'e}zouls}, {Gai}, {Galleti}, {Garabato},
  {Garc{\'\i}a-Sedano}, {Garofalo}, {Garralda}, {Gavel}, {Gavras}, {Gerssen},
  {Geyer}, {Giacobbe}, {Gilmore}, {Girona}, {Giuffrida}, {Glass}, {Gomes},
  {Granvik}, {Gueguen}, {Guerrier}, {Guiraud}, {Guti{\'e}rrez-S{\'a}nchez},
  {Haigron}, {Hatzidimitriou}, {Hauser}, {Haywood}, {Heiter}, {Helmi}, {Heu},
  {Hilger}, {Hobbs}, {Hofmann}, {Holland}, {Huckle}, {Hypki}, {Icardi},
  {Jan{\ss}en}, {Jevardat de Fombelle}, {Jonker}, {Juh{\'a}sz}, {Julbe},
  {Karampelas}, {Kewley}, {Klar}, {Kochoska}, {Kohley}, {Kolenberg},
  {Kontizas}, {Kontizas}, {Koposov}, {Kordopatis}, {Kostrzewa-Rutkowska},
  {Koubsky}, {Lambert}, {Lanza}, {Lasne}, {Lavigne}, {Le Fustec}, {Le
  Poncin-Lafitte}, {Lebreton}, {Leccia}, {Leclerc}, {Lecoeur-Taibi},
  {Lenhardt}, {Leroux}, {Liao}, {Licata}, {Lindstr{\o}m}, {Lister}, {Livanou},
  {Lobel}, {L{\'o}pez}, {Managau}, {Mann}, {Mantelet}, {Marchal}, {Marchant},
  {Marconi}, {Marinoni}, {Marschalk{\'o}}, {Marshall}, {Martino}, {Marton},
  {Mary}, {Massari}, {Matijevi{\v{c}}}, {Mazeh}, {McMillan}, {Messina},
  {Michalik}, {Millar}, {Molina}, {Molinaro}, {Moln{\'a}r}, {Montegriffo},
  {Mor}, {Morbidelli}, {Morel}, {Morris}, {Mulone}, {Muraveva}, {Musella},
  {Nelemans}, {Nicastro}, {Noval}, {O'Mullane}, {Ord{\'e}novic},
  {Ord{\'o}{\~n}ez-Blanco}, {Osborne}, {Pagani}, {Pagano}, {Pailler},
  {Palacin}, {Palaversa}, {Panahi}, {Pawlak}, {Piersimoni}, {Pineau}, {Plachy},
  {Plum}, {Poggio}, {Poujoulet}, {Pr{\v{s}}a}, {Pulone}, {Racero}, {Ragaini},
  {Rambaux}, {Ramos-Lerate}, {Regibo}, {Reyl{\'e}}, {Riclet}, {Ripepi}, {Riva},
  {Rivard}, {Rixon}, {Roegiers}, {Roelens}, {Romero-G{\'o}mez}, {Rowell},
  {Royer}, {Ruiz-Dern}, {Sadowski}, {Sagrist{\`a} Sell{\'e}s}, {Sahlmann},
  {Salgado}, {Salguero}, {Sanna}, {Santana-Ros}, {Sarasso}, {Savietto},
  {Schultheis}, {Sciacca}, {Segol}, {Segovia}, {S{\'e}gransan}, {Shih},
  {Siltala}, {Silva}, {Smart}, {Smith}, {Solano}, {Solitro}, {Sordo}, {Soria
  Nieto}, {Souchay}, {Spagna}, {Spoto}, {Stampa}, {Steele},
  {Steidelm{\"u}ller}, {Stephenson}, {Stoev}, {Suess}, {Surdej}, {Szabados},
  {Szegedi-Elek}, {Tapiador}, {Taris}, {Tauran}, {Taylor}, {Teixeira},
  {Terrett}, {Teyssandier}, {Thuillot}, {Titarenko}, {Torra Clotet}, {Turon},
  {Ulla}, {Utrilla}, {Uzzi}, {Vaillant}, {Valentini}, {Valette}, {van Elteren},
  {Van Hemelryck}, {van Leeuwen}, {Vaschetto}, {Vecchiato}, {Veljanoski},
  {Viala}, {Vicente}, {Vogt}, {von Essen}, {Voss}, {Votruba}, {Voutsinas},
  {Walmsley}, {Weiler}, {Wertz}, {Wevers}, {Wyrzykowski}, {Yoldas},
  {{\v{Z}}erjal}, {Ziaeepour}, {Zorec}, {Zschocke}, {Zucker}, {Zurbach}, \&
  {Zwitter}}]{gaiadr218}
{Gaia Collaboration}, {Brown}, A.~G.~A., {Vallenari}, A., {et~al.} 2018, \aap,
  616, A1, \dodoi{10.1051/0004-6361/201833051}

\bibitem[{{Gehrels}(1986)}]{gehrels1986}
{Gehrels}, N. 1986, \apj, 303, 336, \dodoi{10.1086/164079}

\bibitem[{{Harris}(1996)}]{harr962010}
{Harris}, W.~E. 1996, \aj, 112, 1487, \dodoi{10.1086/118116}

\bibitem[{{Heinke} {et~al.}(2003){Heinke}, {Grindlay}, {Lloyd}, \&
  {Edmonds}}]{heinkeea03}
{Heinke}, C.~O., {Grindlay}, J.~E., {Lloyd}, D.~A., \& {Edmonds}, P.~D. 2003,
  \apj, 588, 452, \dodoi{10.1086/374039}

\bibitem[{{Homan} {et~al.}(2018){Homan}, {van den Berg}, {Heinke}, {Pooley},
  {Degenaar}, {van den Eijnden}, {Bahramian}, {Gendreau}, \&
  {Arzoumanian}}]{homaea18}
{Homan}, J., {van den Berg}, M., {Heinke}, C., {et~al.} 2018, The Astronomer's
  Telegram, 11598, 1

\bibitem[{{Horne}(1985)}]{horn85}
{Horne}, K. 1985, \mnras, 213, 129, \dodoi{10.1093/mnras/213.2.129}

\bibitem[{{in\,'t Zand} {et~al.}(2003{\natexlab{a}}){in\,'t Zand},
  {Strohmayer}, {Markwardt}, \& {Swank}}]{intzea03b}
{in\,'t Zand}, J.~J.~M., {Strohmayer}, T.~E., {Markwardt}, C.~B., \& {Swank},
  J. 2003{\natexlab{a}}, \aap, 409, 659, \dodoi{10.1051/0004-6361:20031042}

\bibitem[{{in\,'t Zand} {et~al.}(2000){in\,'t Zand}, {Bazzano}, {Cocchi},
  {Cornelisse}, {Heise}, {Kuiper}, {Kuulkers}, {Markwardt}, {Muller},
  {Natalucci}, {Smith}, {Strohmayer}, {Ubertini}, \& {Verbunt}}]{intzea00}
{in\,'t Zand}, J.~J.~M., {Bazzano}, A., {Cocchi}, M., {et~al.} 2000, \aap, 355,
  145, \dodoi{10.48550/arXiv.astro-ph/9910107}

\bibitem[{{in\,'t Zand} {et~al.}(2003{\natexlab{b}}){in\,'t Zand}, {Hulleman},
  {Markwardt}, {M{\'e}ndez}, {Kuulkers}, {Cornelisse}, {Heise}, {Strohmayer},
  \& {Verbunt}}]{intzea03}
{in\,'t Zand}, J.~J.~M., {Hulleman}, F., {Markwardt}, C.~B., {et~al.}
  2003{\natexlab{b}}, \aap, 406, 233, \dodoi{10.1051/0004-6361:20030681}

\bibitem[{{Iwai} {et~al.}(2014){Iwai}, {Dotani}, {Ozaki}, {Maeda}, {Mori}, \&
  {Saji}}]{iwai2014}
{Iwai}, M., {Dotani}, T., {Ozaki}, M., {et~al.} 2014, in Suzaku-MAXI 2014:
  Expanding the Frontiers of the X-ray Universe, ed. M.~{Ishida}, R.~{Petre},
  \& K.~{Mitsuda}, 158

\bibitem[{{Kim} {et~al.}(2007){Kim}, {Kim}, {Wilkes}, {Green}, {Kim},
  {Anderson}, {Barkhouse}, {Evans}, {Ivezi{\'c}}, {Karovska}, {Kashyap}, {Lee},
  {Maksym}, {Mossman}, {Silverman}, \& {Tananbaum}}]{kimea07}
{Kim}, M., {Kim}, D.-W., {Wilkes}, B.~J., {et~al.} 2007, \apjs, 169, 401,
  \dodoi{10.1086/511634}

\bibitem[{{Kuulkers} {et~al.}(2003){Kuulkers}, {den Hartog}, {in't Zand},
  {Verbunt}, {Harris}, \& {Cocchi}}]{kuulea03}
{Kuulkers}, E., {den Hartog}, P.~R., {in't Zand}, J.~J.~M., {et~al.} 2003,
  \aap, 399, 663, \dodoi{10.1051/0004-6361:20021781}

\bibitem[{{Markwardt} {et~al.}(2009){Markwardt}, {Altamirano}, {Swank},
  {Strohmayer}, {Linares}, \& {Pereira}}]{markwardt2009}
{Markwardt}, C.~B., {Altamirano}, D., {Swank}, J.~H., {et~al.} 2009, The
  Astronomer's Telegram, 2197, 1

\bibitem[{{Painter} {et~al.}(2024){Painter}, {Stefano}, {Kashyap}, {Soria},
  {Lopez-Miralles}, {Urquhart}, {Steiner}, {Motta}, {Ragozzine}, \&
  {Mori}}]{painter2023}
{Painter}, C., {Stefano}, R.~D., {Kashyap}, V.~L., {et~al.} 2024, \mnras,
  \dodoi{10.1093/mnras/stae164}

\bibitem[{{Pallanca} {et~al.}(2013){Pallanca}, {Dalessandro}, {Ferraro},
  {Lanzoni}, \& {Beccari}}]{pallea13}
{Pallanca}, C., {Dalessandro}, E., {Ferraro}, F.~R., {Lanzoni}, B., \&
  {Beccari}, G. 2013, \apj, 773, 122, \dodoi{10.1088/0004-637X/773/2/122}

\bibitem[{{Panurach} {et~al.}(2021){Panurach}, {Strader}, {Bahramian},
  {Chomiuk}, {Miller-Jones}, {Heinke}, {Maccarone}, {Shishkovsky}, {Sivakoff},
  {Tremou}, {Tudor}, \& {Urquhart}}]{panuea21}
{Panurach}, T., {Strader}, J., {Bahramian}, A., {et~al.} 2021, \apj, 923, 88,
  \dodoi{10.3847/1538-4357/ac2c6b}

\bibitem[{{Pavlinsky} {et~al.}(1994){Pavlinsky}, {Grebenev}, \&
  {Sunyaev}}]{pavlea94}
{Pavlinsky}, M.~N., {Grebenev}, S.~A., \& {Sunyaev}, R.~A. 1994, \apj, 425,
  110, \dodoi{10.1086/173967}

\bibitem[{{Predehl} {et~al.}(1991){Predehl}, {Hasinger}, \&
  {Verbunt}}]{predea91}
{Predehl}, P., {Hasinger}, G., \& {Verbunt}, F. 1991, \aap, 246, L21

\bibitem[{{Revnivtsev} {et~al.}(2002){Revnivtsev}, {Trudolyubov}, \&
  {Borozdin}}]{revnea02}
{Revnivtsev}, M.~G., {Trudolyubov}, S.~P., \& {Borozdin}, K.~N. 2002, Astronomy
  Letters, 28, 237, \dodoi{10.1134/1.1467258}

\bibitem[{{Saji} {et~al.}(2016){Saji}, {Mori}, {Matsumoto}, {Dotani}, {Iwai},
  {Maeda}, {Mitsuishi}, {Ozaki}, \& {Tawara}}]{sajiea16}
{Saji}, S., {Mori}, H., {Matsumoto}, H., {et~al.} 2016, \pasj, 68, S15,
  \dodoi{10.1093/pasj/psw011}

\bibitem[{{Valenti} {et~al.}(2007){Valenti}, {Ferraro}, \&
  {Origlia}}]{valeea07}
{Valenti}, E., {Ferraro}, F.~R., \& {Origlia}, L. 2007, \aj, 133, 1287,
  \dodoi{10.1086/511271}

\bibitem[{{van Paradijs} \& {McClintock}(1994)}]{vanpmccl94}
{van Paradijs}, J., \& {McClintock}, J.~E. 1994, \aap, 290, 133

\bibitem[{{Vats} {et~al.}(2018){Vats}, {Wijnands}, {Parikh}, {Ootes},
  {Degenaar}, \& {Page}}]{vatsea18b}
{Vats}, S., {Wijnands}, R., {Parikh}, A.~S., {et~al.} 2018, \mnras, 477, 2494,
  \dodoi{10.1093/mnras/sty733}

\bibitem[{{Verbunt} \& {Hut}(1987)}]{verbhut87}
{Verbunt}, F., \& {Hut}, P. 1987, in The Origin and Evolution of Neutron Stars,
  ed. D.~J. {Helfand} \& J.~H. {Huang}, Vol. 125, 187

\bibitem[{{Verner} {et~al.}(1996){Verner}, {Ferland}, {Korista}, \&
  {Yakovlev}}]{vern1996}
{Verner}, D.~A., {Ferland}, G.~J., {Korista}, K.~T., \& {Yakovlev}, D.~G. 1996,
  \apj, 465, 487, \dodoi{10.1086/177435}

\bibitem[{{{\v{S}}imon}(2009)}]{simo09}
{{\v{S}}imon}, V. 2009, \na, 14, 443, \dodoi{10.1016/j.newast.2008.12.005}

\bibitem[{{Wijnands} {et~al.}(2003){Wijnands}, {Nowak}, {Miller}, {Homan},
  {Wachter}, \& {Lewin}}]{wijnandsea03}
{Wijnands}, R., {Nowak}, M., {Miller}, J.~M., {et~al.} 2003, \apj, 594, 952,
  \dodoi{10.1086/377122}

\bibitem[{{Wilms} {et~al.}(2000){Wilms}, {Allen}, \& {McCray}}]{wilms2000}
{Wilms}, J., {Allen}, A., \& {McCray}, R. 2000, \apj, 542, 914,
  \dodoi{10.1086/317016}

\end{thebibliography}

\restartappendixnumbering 
\appendix

\section{Relative offsets between Chandra images} \label{app_xalign}

In Table~\ref{app_tab_xoffsets} we summarize the mean relative offsets $\Delta$ between ObsID 23443 
and the other {\em Chandra} imaging observations as derived from aligning sources outside the cluster half-light radius. The offsets are defined as $\Delta{\rm RA} = {\rm RA}_{\rm 23443}-{\rm RA}_{\rm ObsID}$ and $\Delta {\rm Dec}={\rm Dec}_{\rm 23443}-{\rm Dec}_{\rm ObsID}$. The coordinates and {\tt wavdetect} counts of the sources used to compute the offsets are listed in Table~\ref{app_tab_xcoords}. 

The HRC-I observation (ObsID 720) and ObsID 23443 have few sources in common. Moreover, of the four sources outside the cluster half-light radius that are detected in both observations, we excluded one. This source is not located very far from the aimpoint of ObsID 720 ($\sim$3.2\arcmin), yet the events that are associated with it make up an elongated shape. The resulting {\tt wavdetect} coordinates are sensitive to the {\tt wavdetect} parameter settings, yielding differences of $\sim$1\arcsec. Of the remaining three common detections, one has only $6$ counts in ObsID 720. For the alignment of the other ObsIDs, we excluded sources with fewer than 10 counts, but given the small number of sources that ObsIDs 720 and 23443 have in common, we computed the offset with and without this faint source. We find that the results are similar: the members of the source pairs appear to be offset relative to each other by similar amounts with little spread (see the first two rows of Table~\ref{app_tab_xoffsets}).

ObsID 21218 and ObsID 23443 have five sources outside the half-light radius with more than 10 counts in common. The offset of one of these sources ($\Delta$RA=+0.87\arcsec, $\Delta$Dec=$-$0.97\arcsec) deviates much from the typical offsets of the other four (mean offset $\Delta$RA=+0.31\arcsec~with a standard deviation of 0.27\arcsec, and $\Delta$Dec=+0.42\arcsec~with a standard deviation of 0.24\arcsec). We excluded this source for the computation of the offset.

\begin{figure}
\centerline{\includegraphics[width=8.5cm]{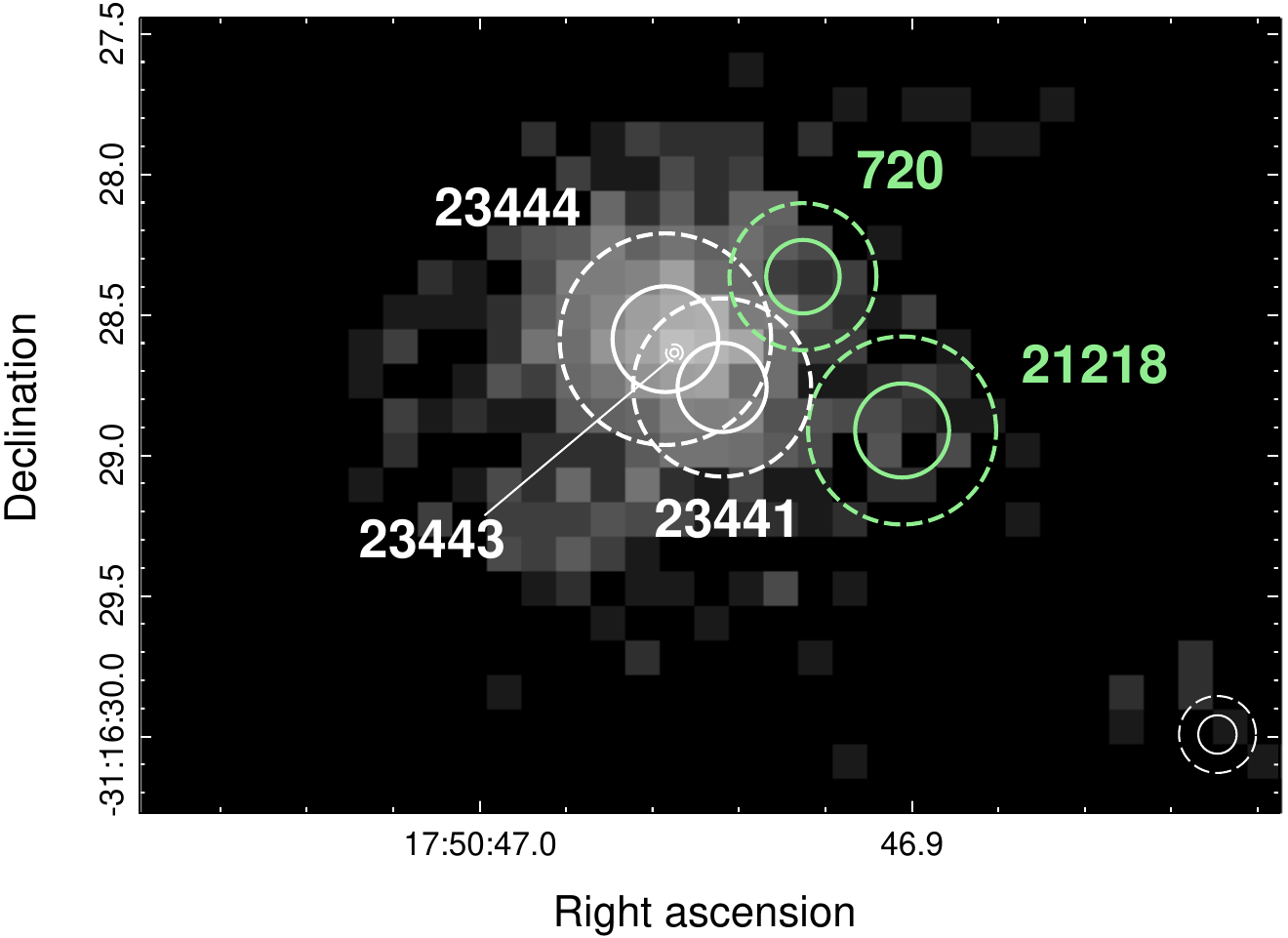}}
\caption{Results of aligning the source lists of the {\em Chandra} imaging observations to the astrometry of ObsID 23443. The 1$\sigma$ and 2$\sigma$ error circles are shown with solid and dashed circles, respectively, where $\sigma$ is computed by adding the {\tt wavdetect} error and the error on the relative astrometric offset in quadrature.  The error circle on the position of X2 measured from ObsID 23443 is very small as it does not include uncertainties in the alignment. The positions shown from ObsIDs 21218 and 23444 are the {\tt wavdetect} positions of the brightest source visible when \grs\ is in eclipse. \label{app_fig1}}
\end{figure}

The resulting aligned positions are shown on top of the ObsID 23443 image of the core of Terzan\,6 in Fig.~\ref{app_fig1}.

\begin{deluxetable}{rrrrrr}[h!]
\caption{Relative astrometric offsets between ObsID 23443 and the other imaging-mode ObsIDs. $N_{\rm align}$ 
is the number of sources used to compute the mean offsets in right ascension ($\Delta$RA) and declination ($\Delta$Dec). The columns e\_${\Delta{\rm RA}}$ and e\_${\Delta{\rm Dec}}$ are the errors on the mean offsets (root-mean-square of the residuals around the mean divided by $\sqrt{N_{\rm align}}$). \label{app_tab_xoffsets}}
\tablehead{ObsID & $N_{\rm align}$ & $\Delta$RA & e\_${\Delta{\rm RA}}$ & $\Delta$Dec & e\_${\Delta{\rm Dec}}$\\ 
                 &                 & (arcsec)   & (arcsec)              & (arcsec)    & (arcsec)            }
\startdata
   720 &  2 &    0.56 & 0.08 & 1.34 & 0.02 \\
   720 &  3 &    0.59 & 0.06 & 1.19 & 0.12 \\
 21218 &  4 &    0.31 & 0.12 & 0.42 & 0.10 \\
 23444 & 15 & $-$0.18 & 0.11 & 0.53 & 0.12 \\
 23441 &  6 &    0.54 & 0.11 & 1.82 & 0.11 \\ 
\enddata
\end{deluxetable}

\begin{deluxetable*}{llllrrlllllrr}
\tabletypesize{\scriptsize}
\tablecaption{Coordinates (RA, Dec) of the sources used to compute the astrometric offsets with respect to ObsID 23443. Errors on the right ascension (e\_RA) and declination (e\_Dec) are the {\tt wavdetect} positional errors in units of arcseconds. Counts (C) and errors on the counts (e\_C) are values reported by {\tt wavdetect} in the 1--7 keV band. \label{app_tab_xcoords}}
\tablehead{RA & e\_RA & Dec & e\_Dec & \multicolumn{1}{c}{C} & \multicolumn{1}{c}{e\_C} & \hspace{1cm} & RA & e\_RA & Dec & e\_Dec & \multicolumn{1}{c}{C} & \multicolumn{1}{c}{e\_C} \\
              & (arcsec) & & (arcsec) & & & & & (arcsec) & & (arcsec) & &}
\startdata
\tableline
\multicolumn{6}{l}{ObsID 23443} & & \multicolumn{6}{l}{ObsID 720} \\
\tableline
17$^{\rm h}$50$^{\rm d}$31\dotsec43  & 0.14 & $-$31$^{\circ}$12\arcmin10\farcs25 & 0.16 & 247 & 16 && 17$^{\rm h}$50$^{\rm d}$31\dotsec38 & 0.35 & $-$31$^{\circ}$12\arcmin11\farcs55 & 0.44 & 34 &  7 \\
17$^{\rm h}$50$^{\rm d}$35\dotsec31  & 0.11 & $-$31$^{\circ}$18\arcmin24\farcs71 & 0.10 & 148 & 12 && 17$^{\rm h}$50$^{\rm d}$35\dotsec28 & 0.22 & $-$31$^{\circ}$18\arcmin26\farcs08 & 0.20 & 32 &  6 \\
17$^{\rm h}$51$^{\rm d}$00\dotsec43  & 0.20 & $-$31$^{\circ}$16\arcmin30\farcs75 & 0.20 &  24 &  5 && 17$^{\rm h}$51$^{\rm d}$00\dotsec38 & 0.32 & $-$31$^{\circ}$16\arcmin31\farcs66 & 0.26 &  6 &  3 \\
\tableline
\multicolumn{6}{l}{ObsID 23443} &  & \multicolumn{6}{l}{ObsID 21218} \\
\tableline
17$^{\rm h}$50$^{\rm d}$31\dotsec43 & 0.14 & $-$31$^{\circ}$12\arcmin10\farcs25 & 0.16 & 247 & 16 && 17$^{\rm h}$50$^{\rm d}$31\dotsec41 & 0.14 & $-$31$^{\circ}$12\farcm10\farcs40 & 0.13 & 60 &  8 \\
17$^{\rm h}$50$^{\rm d}$32\dotsec67 & 0.13 & $-$31$^{\circ}$17\arcmin09\farcs21 & 0.15 &  60 &  8 && 17$^{\rm h}$50$^{\rm d}$32\dotsec64 & 0.10 & $-$31$^{\circ}$17\farcm09\farcs63 & 0.17 & 20 &  5 \\
17$^{\rm h}$50$^{\rm d}$35\dotsec31 & 0.11 & $-$31$^{\circ}$18\arcmin24\farcs71 & 0.10 & 148 & 12 && 17$^{\rm h}$50$^{\rm d}$35\dotsec26 & 0.23 & $-$31$^{\circ}$18\farcm25\farcs45 & 0.22 & 28 &  6 \\
17$^{\rm h}$51$^{\rm d}$00\dotsec82 & 0.42 & $-$31$^{\circ}$12\arcmin47\farcs85 & 0.41 &  30 &  6 && 17$^{\rm h}$51$^{\rm d}$00\dotsec82 & 0.50 & $-$31$^{\circ}$12\farcm48\farcs23 & 0.33 & 53 &  8 \\
\tableline
\multicolumn{6}{l}{ObsID 23443} &  & \multicolumn{6}{l}{ObsID 23444} \\
\tableline
17$^{\rm h}$50$^{\rm d}$32\dotsec67 & 0.13 &$-$31$^{\circ}$17\arcmin09\farcs21  &0.15 & 59  & 8 &&17$^{\rm h}$50$^{\rm d}$32\dotsec70 & 0.10 &$-$31$^{\circ}$17\arcmin09\farcs49 & 0.11 & 74 &  9 \\
17$^{\rm h}$50$^{\rm d}$35\dotsec31 & 0.11 &$-$31$^{\circ}$18\arcmin24\farcs71  &0.10 &148  &12 &&17$^{\rm h}$50$^{\rm d}$35\dotsec31 & 0.05 &$-$31$^{\circ}$18\arcmin25\farcs16 & 0.07 &145 & 13 \\
17$^{\rm h}$50$^{\rm d}$36\dotsec80 & 0.23 &$-$31$^{\circ}$18\arcmin23\farcs63  &0.32 & 34  & 6 &&17$^{\rm h}$50$^{\rm d}$36\dotsec86 & 0.22 &$-$31$^{\circ}$18\arcmin24\farcs01 & 0.23 & 32 &  6 \\
17$^{\rm h}$50$^{\rm d}$39\dotsec02 & 0.28 &$-$31$^{\circ}$20\arcmin35\farcs03  &0.24 & 77  & 9 &&17$^{\rm h}$50$^{\rm d}$39\dotsec06 & 0.15 &$-$31$^{\circ}$20\arcmin34\farcs89 & 0.14 &121 & 12 \\
17$^{\rm h}$50$^{\rm d}$39\dotsec42 & 0.59 &$-$31$^{\circ}$15\arcmin35\farcs14  &0.29 & 11  & 4 &&17$^{\rm h}$50$^{\rm d}$39\dotsec37 & 0.11 &$-$31$^{\circ}$15\arcmin35\farcs94 & 0.20 & 15 &  4 \\
17$^{\rm h}$50$^{\rm d}$39\dotsec82 & 0.13 &$-$31$^{\circ}$17\arcmin23\farcs77  &0.12 & 24  & 5 &&17$^{\rm h}$50$^{\rm d}$39\dotsec84 & 0.13 &$-$31$^{\circ}$17\arcmin24\farcs40 & 0.13 & 22 &  5 \\
17$^{\rm h}$50$^{\rm d}$44\dotsec64 & 0.18 &$-$31$^{\circ}$17\arcmin56\farcs67  &0.13 & 17  & 4 &&17$^{\rm h}$50$^{\rm d}$44\dotsec67 & 0.09 &$-$31$^{\circ}$17\arcmin57\farcs21 & 0.10 & 37 &  6 \\
17$^{\rm h}$50$^{\rm d}$46\dotsec07 & 0.11 &$-$31$^{\circ}$15\arcmin08\farcs97  &0.11 & 10  & 3 &&17$^{\rm h}$50$^{\rm d}$46\dotsec11 & 0.17 &$-$31$^{\circ}$15\arcmin08\farcs86 & 0.17 & 15 &  4 \\
17$^{\rm h}$50$^{\rm d}$53\dotsec02 & 0.13 &$-$31$^{\circ}$15\arcmin38\farcs64  &0.12 & 11  & 4 &&17$^{\rm h}$50$^{\rm d}$53\dotsec06 & 0.22 &$-$31$^{\circ}$15\arcmin38\farcs70 & 0.14 & 11 &  4 \\
17$^{\rm h}$50$^{\rm d}$57\dotsec39 & 0.50 &$-$31$^{\circ}$19\arcmin53\farcs58  &0.27 & 13  & 4 &&17$^{\rm h}$50$^{\rm d}$57\dotsec41 & 0.25 &$-$31$^{\circ}$19\arcmin54\farcs72 & 0.18 & 30 &  6 \\
17$^{\rm h}$50$^{\rm d}$57\dotsec80 & 0.38 &$-$31$^{\circ}$14\arcmin58\farcs75  &0.16 & 10  & 3 &&17$^{\rm h}$50$^{\rm d}$57\dotsec76 & 0.27 &$-$31$^{\circ}$14\arcmin59\farcs90 & 0.20 & 20 &  5 \\
17$^{\rm h}$51$^{\rm d}$00\dotsec43 & 0.20 &$-$31$^{\circ}$16\arcmin30\farcs75  &0.20 & 24  & 5 &&17$^{\rm h}$51$^{\rm d}$00\dotsec43 & 0.20 &$-$31$^{\circ}$16\arcmin31\farcs26 & 0.14 & 30 &  6 \\
17$^{\rm h}$51$^{\rm d}$00\dotsec82 & 0.42 &$-$31$^{\circ}$12\arcmin47\farcs85  &0.41 & 30  & 6 &&17$^{\rm h}$51$^{\rm d}$00\dotsec82 & 0.19 &$-$31$^{\circ}$12\arcmin48\farcs73 & 0.16 &172 & 14 \\
17$^{\rm h}$51$^{\rm d}$02\dotsec23 & 0.45 &$-$31$^{\circ}$18\arcmin37\farcs75  &0.24 & 23  & 5 &&17$^{\rm h}$51$^{\rm d}$02\dotsec27 & 0.52 &$-$31$^{\circ}$18\arcmin37\farcs74 & 0.28 & 12 &  4 \\
17$^{\rm h}$51$^{\rm d}$08\dotsec68 & 0.50 &$-$31$^{\circ}$13\arcmin04\farcs45  &0.35 & 17  & 4 &&17$^{\rm h}$51$^{\rm d}$08\dotsec64 & 0.53 &$-$31$^{\circ}$13\arcmin05\farcs79 & 0.50 & 33 &  7 \\
\tableline
\multicolumn{6}{l}{ObsID 23443} &  & \multicolumn{6}{l}{ObsID 23441} \\
\tableline
   17$^{\rm h}$50$^{\rm d}$31\dotsec43 & 0.14 & $-$31$^{\circ}$12\arcmin10\farcs25 & 0.16 & 247 & 16 && 17$^{\rm h}$50$^{\rm d}$31\dotsec38 & 0.35 & $-$31$^{\circ}$12\arcmin12\farcs36 & 0.24 & 87 & 10 \\
   17$^{\rm h}$50$^{\rm d}$32\dotsec67 & 0.13 & $-$31$^{\circ}$17\arcmin09\farcs21 & 0.15 &  60 &  8 && 17$^{\rm h}$50$^{\rm d}$32\dotsec61 & 0.49 & $-$31$^{\circ}$17\arcmin10\farcs73 & 0.13 & 13 &  4 \\
   17$^{\rm h}$50$^{\rm d}$35\dotsec31 & 0.11 & $-$31$^{\circ}$18\arcmin24\farcs71 & 0.10 & 148 & 12 && 17$^{\rm h}$50$^{\rm d}$35\dotsec24 & 0.11 & $-$31$^{\circ}$18\arcmin26\farcs85 & 0.07 & 60 &  8 \\
   17$^{\rm h}$50$^{\rm d}$36\dotsec80 & 0.23 & $-$31$^{\circ}$18\arcmin23\farcs63 & 0.32 &  34 &  6 && 17$^{\rm h}$50$^{\rm d}$36\dotsec78 & 0.32 & $-$31$^{\circ}$18\arcmin25\farcs35 & 0.22 & 19 &  5 \\
   17$^{\rm h}$50$^{\rm d}$39\dotsec02 & 0.28 & $-$31$^{\circ}$20\arcmin35\farcs03 & 0.24 &  77 &  9 && 17$^{\rm h}$50$^{\rm d}$38\dotsec99 & 0.08 & $-$31$^{\circ}$20\arcmin36\farcs48 & 0.06 &195 & 14 \\
   17$^{\rm h}$50$^{\rm d}$47\dotsec13 & 0.14 & $-$31$^{\circ}$16\arcmin34\farcs76 & 0.09 &  39 &  6 && 17$^{\rm h}$50$^{\rm d}$47\dotsec11 & 0.17 & $-$31$^{\circ}$16\arcmin36\farcs76 & 0.08 & 12 &  4 \\
\enddata

\end{deluxetable*}

\section{Chandra images of the central region of Terzan\,6} \label{app_chandraimages}

\begin{figure*}
\centerline{
\includegraphics[width=17cm,angle=90]{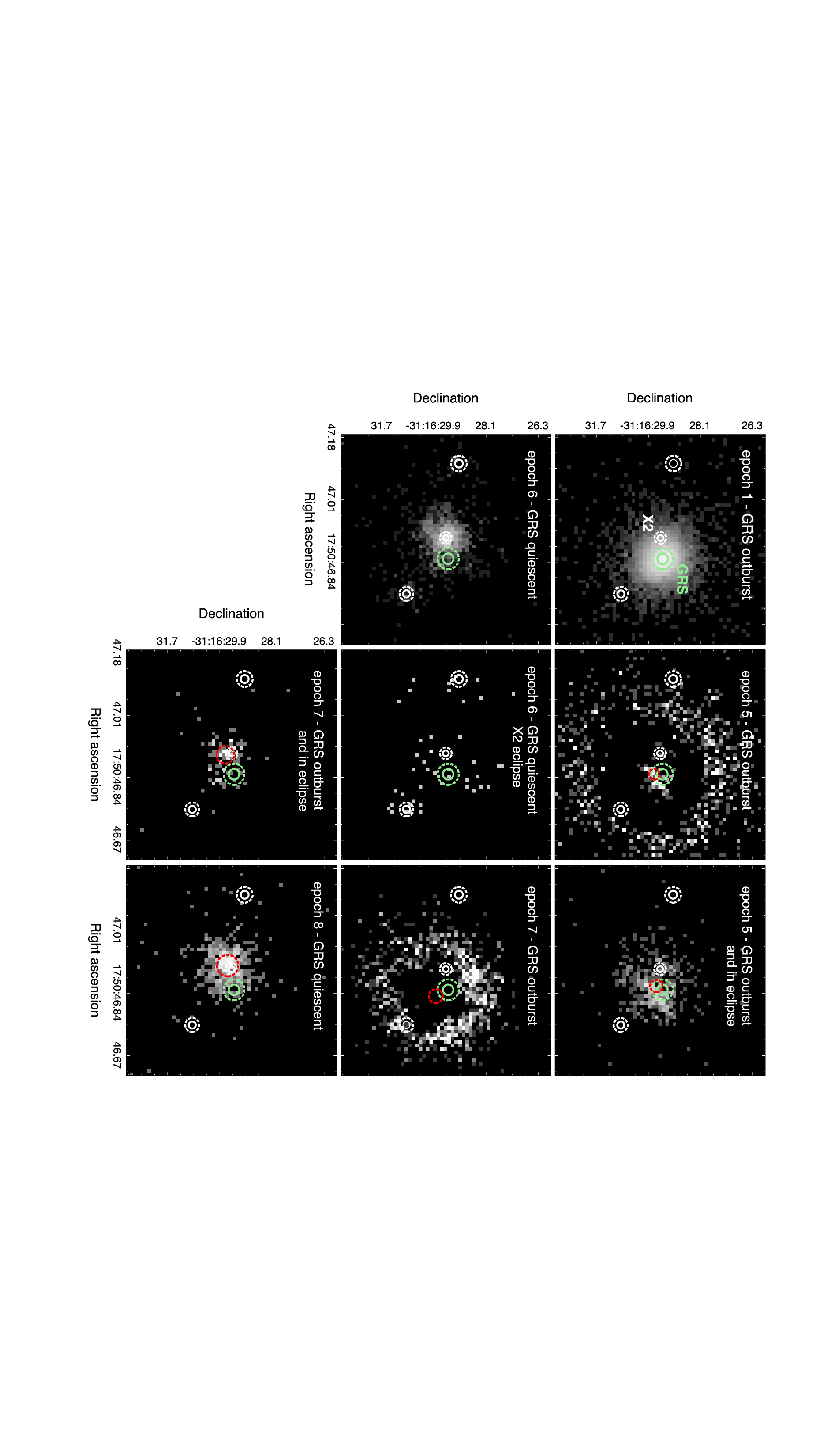}
}
\vspace{-1cm}
\caption{{\em Chandra} images of the core of \ter\ from each ObsID that we analyzed. In each panel we show (in green) the HRC-I detection of \grs\ from epoch 1 when \grs\ was in outburst, and the sources detected in the entire exposure of epoch 6 when \grs\ was in quiescence (in white). The red circles mark the brightest source detected in the {\em Chandra} data shown in the panel. All images are aligned to the {\em HST} astrometry.  The 1- and 2-$\sigma$ errors are shown with solid and dashed circles. The HRC-I image from epoch 1 is in the 0.8--10 keV band. The ACIS-S images from the other epochs are in the 1--7 keV energy band and use 1/4 pixel sub-resolution. North is up, east is to the left.\label{app_fig2}}
\end{figure*}

Each panel in Fig.~\ref{app_fig2} shows the central part of Terzan\,6 in one of the {\em Chandra} images that we analyzed. Like in Fig.~\ref{fig_chandra}, the detection of \grs\ from the HRC-I observation (epoch 1)is shown in green. Sources detected in  the entire exposure of ObsID 23443 (epoch 6) are shown in white. The brightest source detected in ObsIDs 23444 (entire exposure and eclipse), 23441, and 21218 (entire exposure and eclipse) are shown in red (radii are 2 $\sigma$ error circles).

\end{document}